\documentclass[11pt]{article}
\usepackage{jheppub}
\usepackage[utf8x]{inputenc}
\pdfoutput=1
\usepackage{graphicx}
\usepackage{subfigure}
\usepackage{amsmath}
\usepackage{braket}
\usepackage{tensor}
\usepackage{bm}
\usepackage{xspace}
\usepackage{relsize}
\usepackage{placeins}
\usepackage{multirow}
\usepackage{soul}

\def\babar{\mbox{\slshape B\kern-0.1em{\smaller A}\kern-0.1em 
  B\kern-0.1em{\smaller A\kern-0.2em R}}}
\usepackage{lmodern}

\usepackage{color}
\usepackage[usenames,dvipsnames,svgnames,table]{xcolor}
\usepackage{hyperref}
\hypersetup{linktocpage=true}
\usepackage[all]{hypcap}

\pdfoutput=1
\input{header}

\def\carleton{Department of Physics, Carleton University, Ottawa, ON K1S 5B6, Canada}
\def\ucsc{Department of Physics,University of California, Santa Cruz, and Santa Cruz Institute for Particle Physics, Santa Cruz, CA 95064, USA}
\def\uci{Department of Physics and Astronomy, University of California, Irvine Irvine, CA 92697, U.S.A.}
\def\fermi{Fermi National Accelerator Laboratory, Batavia, Illinois 60510, USA}

\preprint{\hfill UCI-HEP-TR-2022-11}

\title{Heavy Neutral Leptons at Beam Dump Experiments of Future Lepton Colliders}

\author[a]{Pierce Giffin,}
\author[a]{Stefania Gori,}
\affiliation[a]{\ucsc}
\author[b,c]{Yu-Dai Tsai,}
\affiliation[b]{\uci}
\affiliation[c]{\fermi}
\author[d]{Douglas Tuckler}
\affiliation[d]{\carleton}

\emailAdd{pgiffin@ucsc.edu}
\emailAdd{sgori@ucsc.edu}
\emailAdd{yudait1@uci.edu}
\emailAdd{dtuckler@physics.carleton.ca}

\date{\today}

\abstract{
A new beam dump experiment that utilizes the beam of future high energy electron-positron colliders could be an excellent avenue to search for dark sector particles due to its unprecedented high energy and intensity. We consider heavy neutral leptons (HNLs) as a specific example to demonstrate the sensitivity of searches for dark sector particles at future electron-positron collider beam dump experiments. This includes the study of the reach at the International Linear Collider (ILC), the Cool Copper Collider ($\rm C^3$), and the Compact Linear Collider (CLIC). We comprehensively examine the HNL production and detector acceptance at these electron beam dump experiments.
We show that these experiments will probe a large range of HNL parameter space, not yet probed by past experiments. These experiments will be complementary to 
other proposed experiments such as proton beam dump experiments, neutrino experiments, and LHC auxiliary detectors. Our study also motivates a more detailed analysis of heavy meson productions in high-energy electron-nucleon collisions in thick targets.}

\begin{document}
\maketitle
\flushbottom

\section{Introduction}

Heavy neutral leptons (HNLs) are well-motivated extensions of the Standard Model (SM) \cite{Minkowski:1977sc,GellMann:1980vs,Yanagida:1979as,Glashow:1979nm,Mohapatra:1979ia,Schechter:1980gr,Foot:1988aq}. The observation of neutrino oscillations is evidence for non-zero neutrino masses. A straightforward
way of generating them is to add to the SM right-handed gauge-singlet fermions, $N_i$, called HNLs.
In addition, HNLs naturally appear in many other extension of the SM, since they can be connected to the matter-antimatter asymmetry problem and to the nature of Dark Matter (see, e.g., \cite{Asaka:2005an,Asaka:2005pn,Bondarenko:2018ptm,Abdullahi:2022jlv}.) 
A rich experimental program has been conducted in recent years to probe HNL parameter space both at high intensity and at high energy experiments. This includes neutrino experiments (e.g., CHARM~\cite{Orloff:2002de}, PS191~\cite{Bernardi:1985ny,Bernardi:1987ek}, NOMAD~\cite{NOMAD:2001eyx},  T2K~\cite{T2K:2019jwa,Arguelles:2021dqn}, MicroBooNE~\cite{MicroBooNE:2019izn,Kelly:2021xbv},  ArgoNeuT~\cite{ArgoNeuT:2021clc}, the Fermilab SBN detectors~\cite{Ballett:2016opr}, and the DUNE near detector complex~\cite{Ballett:2019bgd,Berryman:2019dme,Coloma:2020lgy}), proposed auxiliary LHC detectors that target long-lived particles (e.g. FASER~\cite{Kling:2018wct}, MATHUSLA~\cite{MATHUSLA:2018bqv,Curtin:2018mvb,MATHUSLA:2019qpy,MATHUSLA:2020uve}, CODEX-b~\cite{Gligorov:2017nwh,Aielli:2019ivi}), and proton fixed-target beam dump experiments (e.g. DarkQuest~\cite{Batell:2020vqn}, SHiP~\cite{Gninenko:2012anz,Bonivento:2013jag,Alekhin:2015byh,SHiP:2018xqw,Gorbunov:2020rjx}, and NA62~\cite{Drewes:2018gkc,Chun:2019nwi}). In addition, there are searches at high energy proton-proton colliders such as the LHC, including its ATLAS \cite{ATLAS:2019kpx} and LHCb detector \cite{LHCb:2020wxx}, proposals for future searches at the High-Luminosity LHC (HL-LHC) \cite{Cheung:2020buy}, and at existing and future $e^+ e^-$ colliders, including Belle \cite{Belle:2013ytx}, Belle II~\cite{Dreyer:2021aqd}, and the Future Circular Collider (FCC) \cite{Blondel:2014bra}. HNLs can produced in several different processes at these experiments. In particular, HNL are produced from SM meson decays at neutrino and proton beam dump experiments, from Drell-Yan processes mediated by the weak interactions and from $Z$ or $W$ decays at the LHC and $e^+e^-$ colliders.

Here, we propose to use the beam dump of future electron-positron colliders, such as the International Linear Collider (ILC), the Cool Copper Collider (C$^3$) \cite{Bai:2021rdg}, and FCC-ee \cite{Blondel:2014bra,Chrzaszcz:2020emg}, to search for HNLs. As we will show, adding instruments after the electron beam dump of these experiments could lead to the exploration of new HNL parameter space. While the main goal of these next-generation $e^+e^-$ machines is to use the main collision to perform precision studies of SM particles (such as the Higgs boson and the top quark) and to produce light New Physics (NP) particles, they could also perform searches for the production of new dark sector (i.e. gauge singlet) states produced in the electron beam dump.
More specifically, once the electrons and positrons pass the collision point, they will be discarded in the beam dump. Thanks to the large number of electrons on target, this has the potential to produce a large amount of light dark sector particles from the interactions of the electrons or positrons with the beam dump. These dark sector particles can subsequently decay into visible SM particles, that could be detected by a suitable apparatus installed behind the beam dump. It has been shown that an ILC beam dump experiment can probe currently unexplored parameter space of many well-motivated Beyond the Standard Model (BSM) theories (dark photons~\cite{Kanemura:2015cxa,Asai:2021ehn}, axion-like particles (ALPs) and light scalar particles~\cite{Sakaki:2020mqb,Asai:2021ehn}, and leptophilic Gauge Bosons~\cite{Asai:2021xtg}). 

In this paper, we discuss the possibility of searching for HNLs at  high-energy electron beam dump experiments where HNLs can be produced in charged-current electron-proton scattering and in the decays of mesons produced after the electrons collide with the beam dump. Once produced the HNLs can travel a macroscopic distance before decaying into SM particles. These particles can be detected if a suitable detector is installed behind the beam dump. Depending on the specific setup of the experiment, this type of signatures can have a very low background. Hence, these searches will be complementary to the HNL searches that can be performed using the main $e^+e^-$ collider~\cite{L3:2001xsz,Blondel:2014bra,Antusch:2015mia,Antusch:2016ejd,Alimena:2022hfr}.
Our study is also relevant for future electron-ion colliders~\cite{Batell:2022ubw} as they share similar production channels, especially the charged-current scattering for electron-mixed HNL discussed in our work.

Future electron collider beam dump experiments will provide exciting opportunities due to their unprecedented high energy and high intensity beams. Compared to existing and near-future proton beam dump experiments, electron collider beam dumps can reach comparable or higher center-of-mass energies and can thus produce heavier dark sector particles.
Far forward experiments at hadron colliders have higher energies \cite{Anchordoqui:2021ghd,Feng:2022inv}, but they have lower intensity compared to electron collider beam dumps due to their geometric acceptance and lower luminosities.
The main beam dumps of hadron collider could also be interesting for probing dark sector particles, but they suffer from significant radiation that can easily overwhelm the detectors if the detectors are directly placed downstream of the dump \cite{Kelly:2021jgj}.

This paper is organized as follows. In section~\ref{sec:ILC}, we give an overview of the beam dump experiments that could be performed at future $e^+e^-$ colliders, including, ILC, C$^3$, and FCC-ee. In section~\ref{sec:HNL}, we discuss the phenomenology of HNLs. This is followed by the prospects for HNL searches at electron collider beam dump experiments in section~\ref{sec:reach}. We conclude in section~\ref{sec:conclusion}. In the Appendix, we report the details of our simulations for meson production at the beam dump. The meson production rate will be the main source of uncertainty in the calculation of the reach of the HNL parameter space.

\section{Beam dump Experiments at Future $e^+e^-$ Colliders}\label{sec:ILC}

We begin by describing the experimental setup under consideration, following the proposal for a beam dump experiment at the ILC in \cite{Kanemura:2015cxa,Sakaki:2020mqb}. We highlight that similar setups could be implemented at other future $e^+e^-$ linear colliders like C$^3$ and CLIC, as well as at future $e^+e^-$ circular colliders like FCC-ee and CEPC.


\subsection{ILC beam dump Configuration}\label{Sec:Setup}

\begin{figure}
    \centering
    \includegraphics[width=1\textwidth]{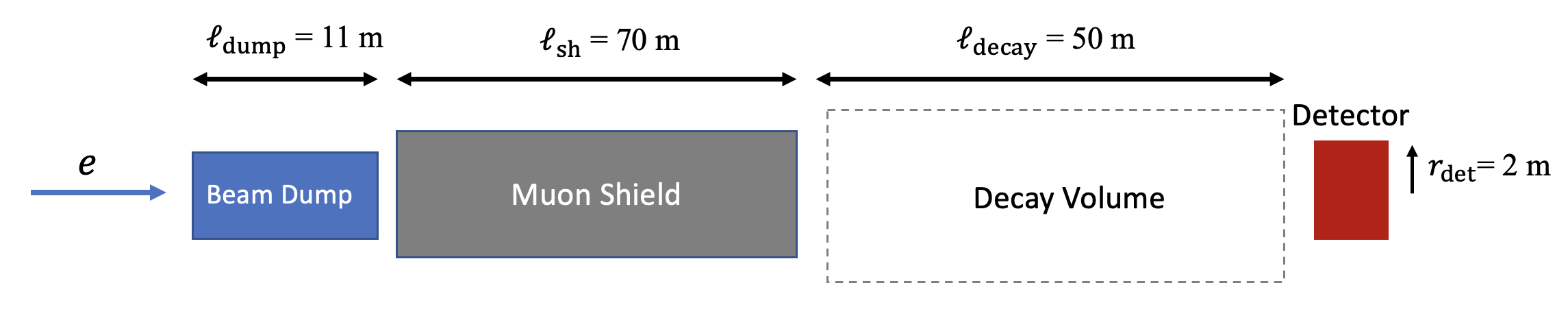}
    \caption{Schematic layout of the beam dump experiment for the ILC. A similar experiment can be designed for other $e^+e^-$ colliders as $\rm C^3$, CLIC, FCC-ee, and CEPC. 
   \label{fig:configuration}
   }
\end{figure}

The International Linear Collider (ILC) experiment is a proposed future collider that uses high-energy collisions of electron ($e^{−}$) and positron ($e^{+}$) beams. The ILC is planned to initially run with a center-of-mass energy of 250 GeV, and will then be upgraded to run at 350 GeV, 500 GeV, and 1 TeV \cite{ILCInternationalDevelopmentTeam:2022izu}. In addition to its central collider, the ILC can host a number of additional detectors, including detectors for beam dump experiments. 

The main dumps accept the full power of the ILC beam. We consider the configuration described in \cite{Sakaki:2020mqb} (see also section 11.2 of the Snowmass report \cite{ILCInternationalDevelopmentTeam:2022izu}).
The length of the main dump which uses water as the absorber is $\ell_{dump} = 11\;\rm m$.
This is followed by a muon shield made of lead and concrete of length $\ell_{sh} = 70\;\rm m$ along the beam direction, and by a decay volume of length $\ell_{decay} = 50\;\rm m$. The detector is assumed to be cylindrical with a radius  $r_{det}=2\;\rm m$ and its axis aligned with the beam axis. We will make the simple assumption that the detector has a 100$\%$ detection efficiency, as long as SM particles are in geometric acceptance. A schematic layout of the experiment is shown in figure~\ref{fig:configuration}.

In the following sections, we will focus and compare the physics opportunities that arise using the 125 GeV ($\sqrt s=250$ GeV) and 500 GeV ($\sqrt s=1$ TeV) ILC electron beams, which we will denote by ``ILC-250'' and ``ILC-1000'', respectively. For both stages, we will consider the initial ILC-250 configuration which can deliver $4\times 10^{21}$ electrons-on-target (EOT) per year (see the ILC beam parameters discussed in table 4.1 of \cite{ILCInternationalDevelopmentTeam:2022izu}).
After the first ILC stage, the number of EOT/year can be increased up to $10^{22}$ EOT/year in the most optimistic scenario~\cite{ILCInternationalDevelopmentTeam:2022izu}.
Thanks to these large luminosities, a sizable flux of mesons will be created. Most of these mesons are stopped, absorbed, or decay in the dump. Particles surviving the dump (e.g. muons) are shielded by the muon shield. New physics long-lived particles, like heavy neutral leptons can be produced from meson decays in the dump, travel a macroscopic length, and decay in the decay volume. The visible decay products can be searched for by the detector placed at the end of the decay volume.

\setlength{\tabcolsep}{0.65em}
{
\renewcommand{\arraystretch}{1.25} 
\begin{table*}[t] 
\begin{tabular}{| c | c | c | c | c | }
\hline Collider-$\sqrt{s}$ [GeV]
    & ILC-250/1000 & C$^3$-250 &  C$^3$-3000 & CLIC-3000 \\ \hline
   Bunches/Train & 1312 & 133 & 75 & 312\\
   Train Rep. Rate [Hz] & 5 & 120 & 120 & 50\\
   Bunch Charge [nC] & 3.2 & 1 & 1 &0.6 \\
Effective Luminosity [cm$^{-2}$ s$^{-1}$] & $1.6\times 10^{39}$ & $1.2\times 10^{39}$ & $6.9\times 10^{38}$ & $6.9\times 10^{38}$ 
\\ \hline
   EOT/Year& $4.1\times 10^{21}$ & $3.1\times 10^{21}$& $1.8 \times 10^{21}$ & $1.8\times 10^{21}$  \\
    \hline
  \end{tabular}
\caption{
Beam parameters and total number of electrons-on-target (EOT) per year for different $e^+ e^-$ colliders assuming one year of operation. The numbers for the beam parameters are taken from table 4.1 of \cite{ILCInternationalDevelopmentTeam:2022izu} for ILC, table 3 of \cite{Bai:2021rdg} for C$^3$, and table 4.1 of \cite{Aicheler:2018arh} for CLIC. The ``effective luminosity'', $\mathcal{L}_\text{eff}$, refers to the instantaneous luminosity of electron-nucleus collisions in the first radiation length of the dump (and not to the instantaneous luminosity of the corresponding primary $e^+e^-$ collision).
}\label{tab:BeamParams}
\end{table*}
}

\subsection{Beam dump Experiments at other $e^+e^-$ Colliders: C$^3$, CLIC, FCC-ee, CEPC}

In addition to the ILC, several next-generation $e^+e^-$ colliders have been proposed for precision SM and BSM physics studies, such as the Cool Copper Collider (C$^3$), the Compact Linear Collider (CLIC), the Future Circular Collider (FCC), and the Circular Electron Positron Collider (CEPC). In this section, we discuss how these experiments compare to the ILC. 

C$^3$ is a recently proposed $e^+e^-$ linear collider~\cite{Bai:2021rdg,Dasu:2022nux,Nanni:2022oha}. C$^3$ will initially run at a 250 GeV center-of-mass energy and has the possibility of an upgrade to 3 TeV. We denote these two runs by ``C$^3$-250'' and ``C$^3$-3000'', respectively. The beam configurations of these two stages differ only in the number of bunches-per-train. The numbers of EOT/year are $3.1\times 10^{21}$ and $1.8 \times 10^{21}$, at C$^3$-250 and C$^3$-3000, respectively. These numbers are similar to the number of EOT at an ILC beam dump experiment (see table \ref{tab:BeamParams}).

CLIC is a proposed multi-TeV linear collider designed to operate in three stages with center-of-mass energies of 380 GeV, 1.5 TeV, and 3 TeV. Using the beam parameters reported in \cite{Aicheler:2018arh}, we find a luminosity of $1.8\times 10^{21}$ EOT/year for the 3 TeV stage, denoted ``CLIC-3000'' in table~\ref{tab:BeamParams}. The luminosities for the other stages are similar. 

FCC is a proposed next-generation circular collider, which includes a high luminosity $e^+e^-$ storage ring collider (FCC-ee), with center-of-mass energies between 90 and 350 GeV \cite{FCC:2018evy}. The planned collider luminosity will be higher than at the ILC. However, given the nature of circular colliders, the beams are reused and dumped much less frequently, yielding a lower number of EOT per year. Approximately $10^{17}$ electrons will be dumped per year at FCC-ee \cite{Kanemura:2015cxa}.
Similarly, CEPC is an electron-positron circular collider that is planned to run with a center-of-mass energy of 240 GeV. Since CEPC is a circular collider, it will also lead to a much smaller number of EOT per year.
For this reason, in this paper, we will focus on the reach of the HNL parameter space at linear $e^+e^-$ colliders, focusing on the ILC-250, ILC-1000, and C$^3$/CLIC-3000 experiments.\footnote{As shown in table \ref{tab:BeamParams}, from the point of view of the energy and the luminosity, CLIC running at $\sqrt{s}$ = 3 TeV is identical to C$^3$-3000, and C$^3$-250 is identical to ILC-250.} We summarize the parameters of these experiments in table~\ref{tab:BeamParams}.

In the literature, there has not been any discussion of beam dump experiments at C$^3$ or CLIC. In the following sections, we will assume that the main beam dump of CLIC and C$^3$ is based on the ILC beam dump design \cite{Aicheler:2012bya, Aicheler:2018arh}.
In section \ref{sec:mod}, we will discuss how different beam dump configurations will affect the reach of the HNL parameter space.

We hope our study will provide strong motivation to consider beam dump experiments at these future electron-positron colliders, and will inform the optimal beam dump configuration to maximize the reach on the HNL parameter space, and, more in general, on long-lived dark sector particles produced from meson decays.

\subsection{Meson Production at high-energy electron beam dumps}

Because of the high energy of the electron beam and the large number of electrons on target, mesons will be substantially produced in the beam dump. To determine the production rate of a given meson, we use \texttt{Pythia 8.3} \cite{Sjostrand:2006za,Sjostrand:2014zea,Bierlich:2022pfr} to simulate 125 GeV, 500 GeV, and 1.5 TeV electron beams striking a fixed target using the \texttt{SoftQCD:all} event class. 
In Appendix~\ref{app:sim}, we collect the details of our simulations. We only consider the mesons produced in the first radiation length of the dump. This will lead to a conservative estimate of the number of mesons produced. Unfortunately, there is no past data on meson production in electron-nucleus collisions at these energies that can be used to normalize the rates. For this reason, the number of mesons produced will be the main source of uncertainty in the estimate of the reach on the HNL parameter space.

Pions and kaons have large production rates in electron-nucleus collisions.
Given their relatively long proper lifetime, they have a decay length that is much larger than the characteristic interaction length in the beam dump material. As a result, a sizable fraction of these mesons can be absorbed by the beam dump before they decay into HNLs. To take this into account, we estimate the number of mesons that decay before the first interaction length in liquid water as \cite{Batell:2020vqn}
\begin{equation}\label{eq:Nmeson}
N_M \approx N_{\rm{EOT}}~ n_M \Gamma_M \langle \gamma_M^{-1}\rangle \lambda_M\,,
\end{equation}
where $N_{\rm{EOT}}$ is the number of electrons on target, $n_M$ is the number of mesons produced in the first radiation length of the dump per electron on target (see eq.~\ref{eq:nMeson} and table~\ref{tab:mesonfrac}), $\langle \gamma_M^{-1}\rangle$ is the mean inverse Lorentz boost, and $\lambda_M$ is the meson interaction length in liquid water. We assume that the interaction length of pions and kaons in liquid water are similar and use $\lambda_M \approx 115$ cm \cite{PDG:Atomic}. 

$D$ and $B$ mesons are short-lived, and, therefore, the number of these mesons responsible of HNL production is simply given by $N_M = N_{\rm{EOT}}~ n_M$.  A large number of $\tau$ leptons are also produced, mainly from $D_s$ meson decays, and the total number is given by $N_\tau = N_{\rm{EOT}}~ n_{D_s} \times \text{BR}(D_s \to \tau \nu_\tau)$.

In table \ref{tab:Nmesons}, we summarize the number of mesons and $\tau$ leptons produced at the several electron beam dump experiments per year of running. In the case of light mesons, the ILC-1000 numbers are slightly larger than the corresponding ones at C$^3$/CLIC-3000, despite its lower energy.
This is a reflection of the lower number of EOT/year of C$^3$/CLIC-3000 compared to the ILC, which compensates for the larger production of mesons per EOT (see table~\ref{tab:mesonfrac}). The kinematical suppression of $B^\pm$ production at ILC-1000 is more relevant. In fact, C$^3$/CLIC-3000 will be able to produce a larger sample of $B^\pm$ mesons, thanks to its higher center-of-mass energy ($\sqrt s \sim 55$ GeV vs. $\sqrt s\sim 30$ GeV of the ILC-1000).

 \setlength{\tabcolsep}{0.65em}
{\renewcommand{\arraystretch}{1.25} 
\begin{table*}[t] 
\centering
\begin{tabular}{| c | c | c | c | c | c |}
\hline
    & ILC-250 & ILC-1000 & C$^3$-250 &  C$^3$/CLIC-3000  \\ \hline
    $\pi^\pm$  & $1.5 \times 10^{16}$& $6.0\times 10^{16}$& $1.2 \times 10^{16}$& $4.8 \times 10^{16}$ \\
    $K^\pm$  & $3.9 \times 10^{15}$& $1.5 \times 10^{16}$ & $3.0 \times 10^{15}$ & $1.2 \times 10^{16}$ \\
    $K_L^0$  & $3.8 \times 10^{14}$&  $1.5\times 10^{15}$& $2.9 \times 10^{14}$ & $1.2 \times 10^{15}$ \\
    $D^\pm$  & $1.4 \times 10^{14}$&  $7.8 \times 10^{14}$& $1.0 \times 10^{14}$& $8.6 \times 10^{14}$ \\
   $D_s^\pm$  & $3.8 \times 10^{13}$&  $2.5 \times 10^{14}$& $2.9 \times 10^{13}$& $2.6 \times 10^{14}$ \\
    $B^\pm$  & $1.0\times10^{11}$& $2.2\times10^{12}$ & $7.5 \times 10^{10}$ & $3.4 \times 10^{12}$ \\
    $\tau^\pm$  &$2.2 \times 10^{12}$ & $1.4 \times 10^{13}$ &  $1.5 \times 10^{12}$& $1.5 \times 10^{13}$\\
    \hline
  \end{tabular}
\caption{Total number of mesons and $\tau$ leptons produced per year at the several beam dump experiments, summing over both charged states. For pions and kaons we show the number of mesons that decay before the first interaction length, not the total number of produced mesons.
} \label{tab:Nmesons}
\end{table*}}

\section{Heavy Neutral Leptons}\label{sec:HNL}

HNLs can interact with the SM lepton doublet via the neutrino portal operator
\begin{equation}\label{eq:HNLLagrangian}
-\mathcal{L}_\text{HNL} \supset Y_{ij} \hat L_i H N_j +{\rm{h.c.}}\,, \end{equation}
where $H$ is the SM Higgs doublet, $\hat L_i = (\nu_i, \ell_i)^T$ is the SM lepton doublet of flavor $i$, and $Y_{ij}$ are the $ij$ elements of the Yukawa matrix. After electroweak symmetry breaking, the HNLs will mix with the SM neutrinos with a $3\times 3$ mixing matrix $U_{ij}$. These mixings induce the coupling of HNLs to electroweak gauge bosons $W^\pm,Z$ given by
\begin{equation}\label{eq:Lweak}
\mathcal{L} \supset  \frac{g_2}{\sqrt{2}} U_{ij} W^-_\mu \ell_i^\dagger \bar\sigma^\mu N_j+\frac{g_2}{2\cos\theta_W} U_{ij} Z_\mu \nu_i^\dagger \bar\sigma^\mu N_j +{\rm{h.c.}}\,,
\end{equation}
where $g_2$ is the $SU(2)_L$ gauge coupling constant and $\theta_W$ is the Weinberg angle.

In our analysis, we will take a phenomenological approach, often adopted in the literature, and assume the existence of one HNL that mixes with only one SM neutrino flavor. We will separately study the cases in which the HNL mixes with the SM electron neutrino with a mixing angle denoted by $U_e$, with the SM muon neutrino with $U_\mu$, and with the SM $\tau$ neutrino with $U_\tau$. The phenomenology is completely determined once the HNL mass, $m_N$, and mixing angle, $U_\alpha$, are fixed.


\subsection{\boldmath HNL Production at an Electron Beam Dump Experiment}\label{subsec:HNLprod}

The mixings generated by the Lagrangian in eq.~(\ref{eq:HNLLagrangian}) allow for HNLs to be produced in any process where a SM neutrino is produced. This allows for the production of HNLs at an electron beam dump experiment in: (1) charged-current electron-proton scattering; (2) two- or three-body (semi-) leptonic decays of mesons (e.g. $M^\pm \to \ell^\pm N$, $M^\pm \to M^{\prime 0} \ell^\pm N$ with $M, M^\prime$ being different mesons); (3) secondary production from the decays of $\tau^\pm$ produced from the decays of heavy mesons such as $D^\pm_s$.

The first process only happens to leading order in the case of electron-mixed HNLs. The corresponding production cross-section is given by (see also \cite{Buchmuller:1991tu}, where we neglect the contributions from heavy quark PDFs)
\begin{equation}
\sigma(ep \to N n) \approx \frac{G^2_F M^2_W}{2\pi}|V_{ud}|^2 |U_e|^2\int_{m^2_N/s}^1 dx \frac{(\hat{s}-m^2_N)^2}{\hat{s}(\hat{s}-m^2_N + M^2_W)}f_u(x,q^2)\,,
\end{equation}
where $p$ and $n$ denote a proton and a neutron, $\hat{s} = x s$ with $\sqrt{s} = \sqrt{2m_p E_e}$ being the center-of-mass energy and $x$ the fraction of the proton momentum carried by the inital state quark. The function $f_u(x,q^2)$ is the up-quark parton distribution function. We have numerically checked that \texttt{MadGraph5\_aMC@NLO} \cite{Alwall:2014hca} gives approximately the same cross section.

To determine the number of electron-mixed HNLs produced directly through the charged-current exchange, we estimate the effective instantaneous luminosity $\mathcal{L}_\text{eff}$ of the electron-proton system in the first radiation length of the beam dump as
\begin{equation}\label{eq:effLumi}
    \mathcal{L}_\text{eff} \simeq
\frac{Z (X_0/g)N_A}{A} \rm\; EOT_s\,,
\end{equation}
where $A(Z)$ is the atomic mass (number) of the target material, $X_0$ is the radiation length ($36.08\;\rm g/cm^2$ for liquid water), $N_A$ is the Avogadro's number, and EOT$_s$ is the number of electrons-on-target per second. The values of $\mathcal{L}_\text{eff}$ for the different experiments are reported in table~\ref{tab:BeamParams}.  

\begin{figure}
    \centering
    \includegraphics[width=0.475\textwidth]{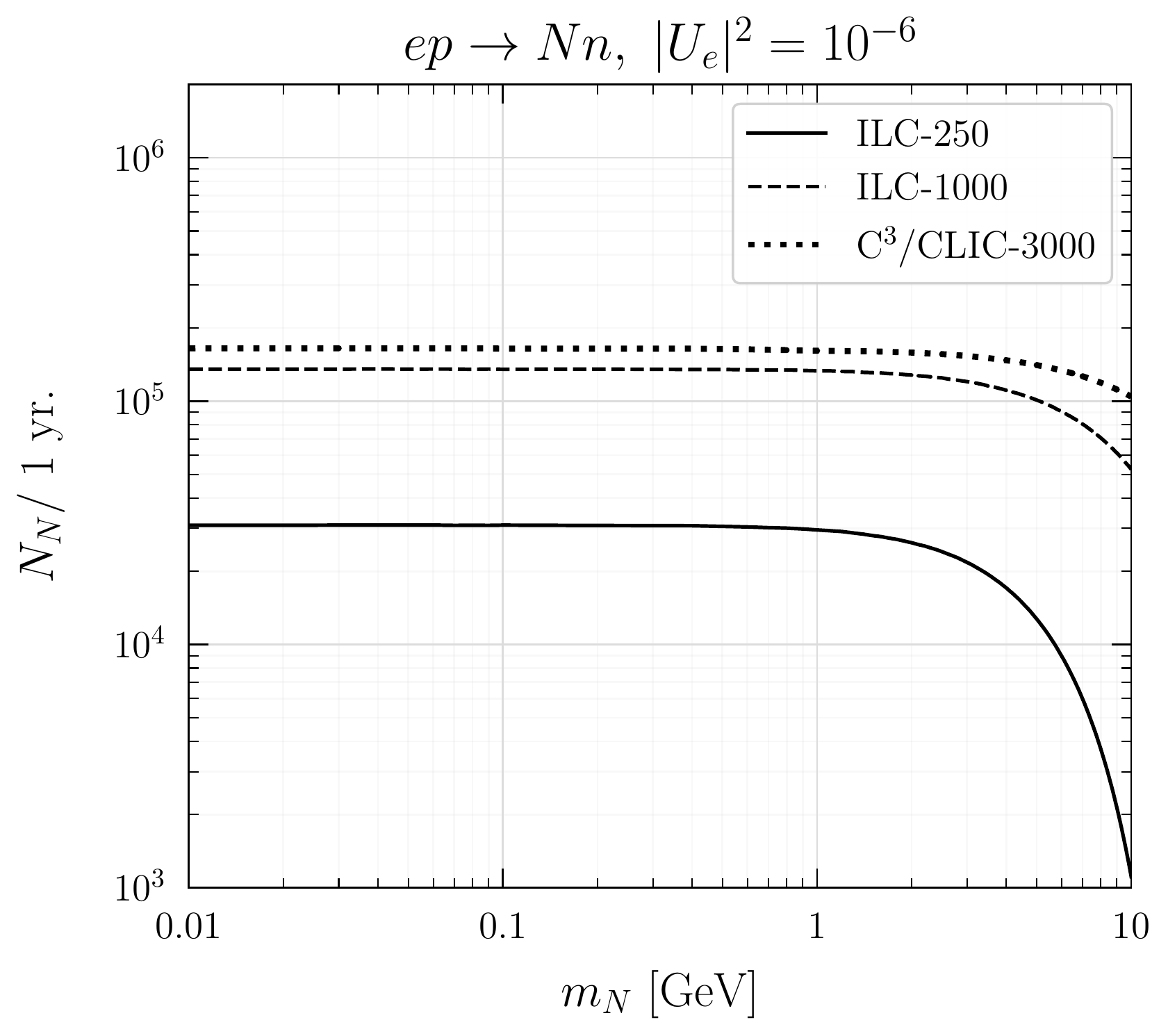}
     \includegraphics[width=0.49\textwidth]{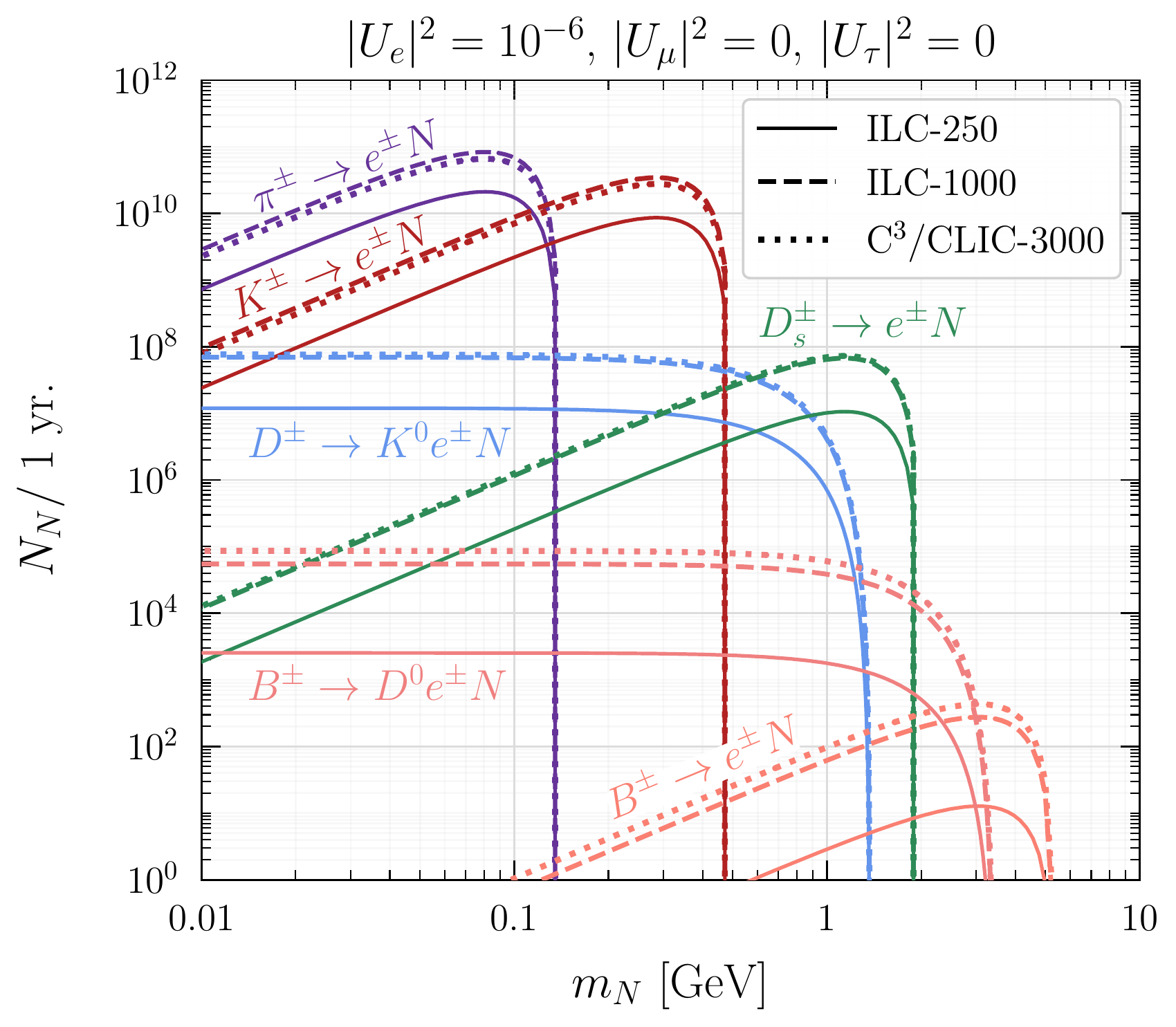}\vspace{0.5cm}
       \includegraphics[width=0.475\textwidth]{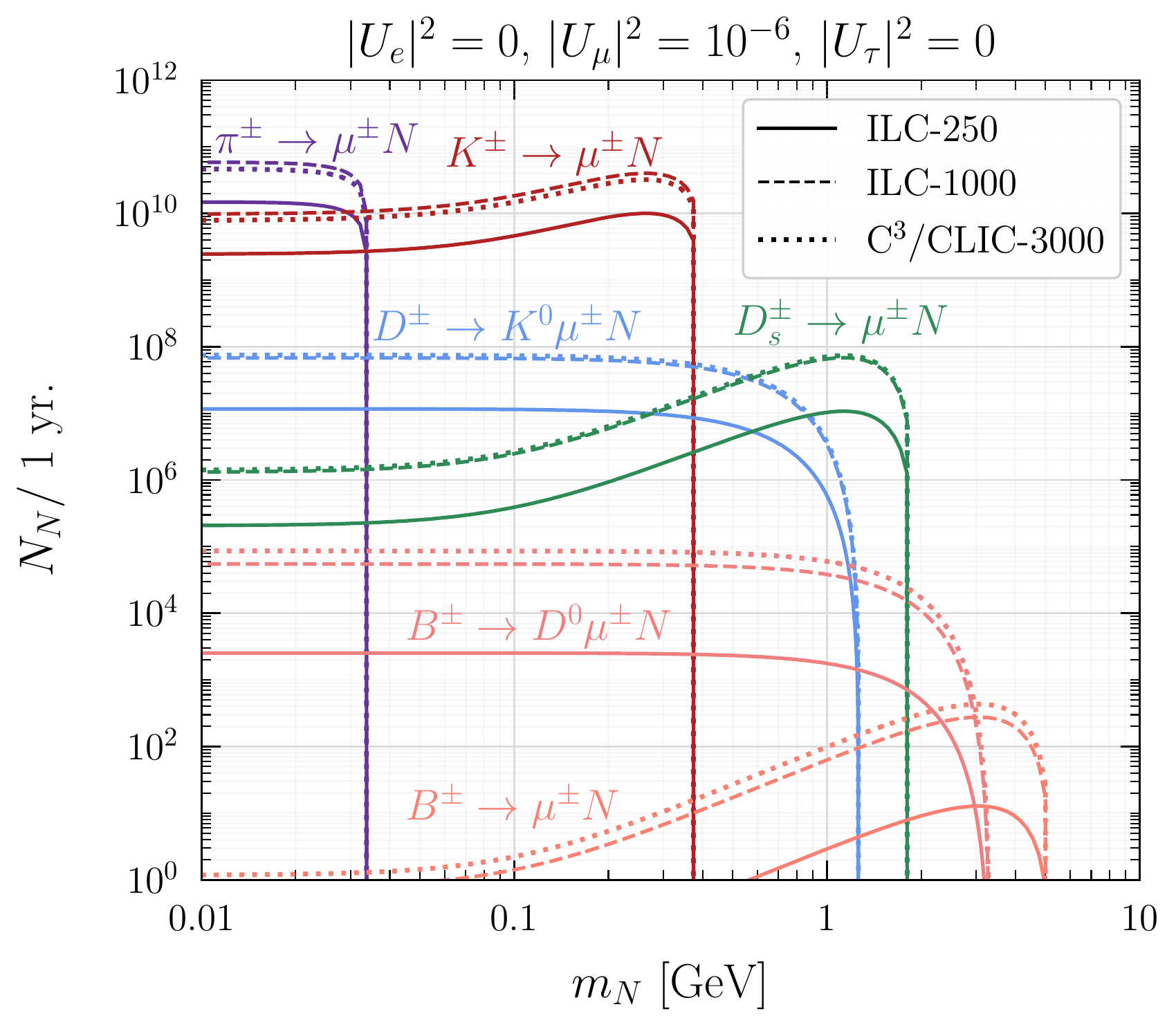}
         \includegraphics[width=0.475\textwidth]{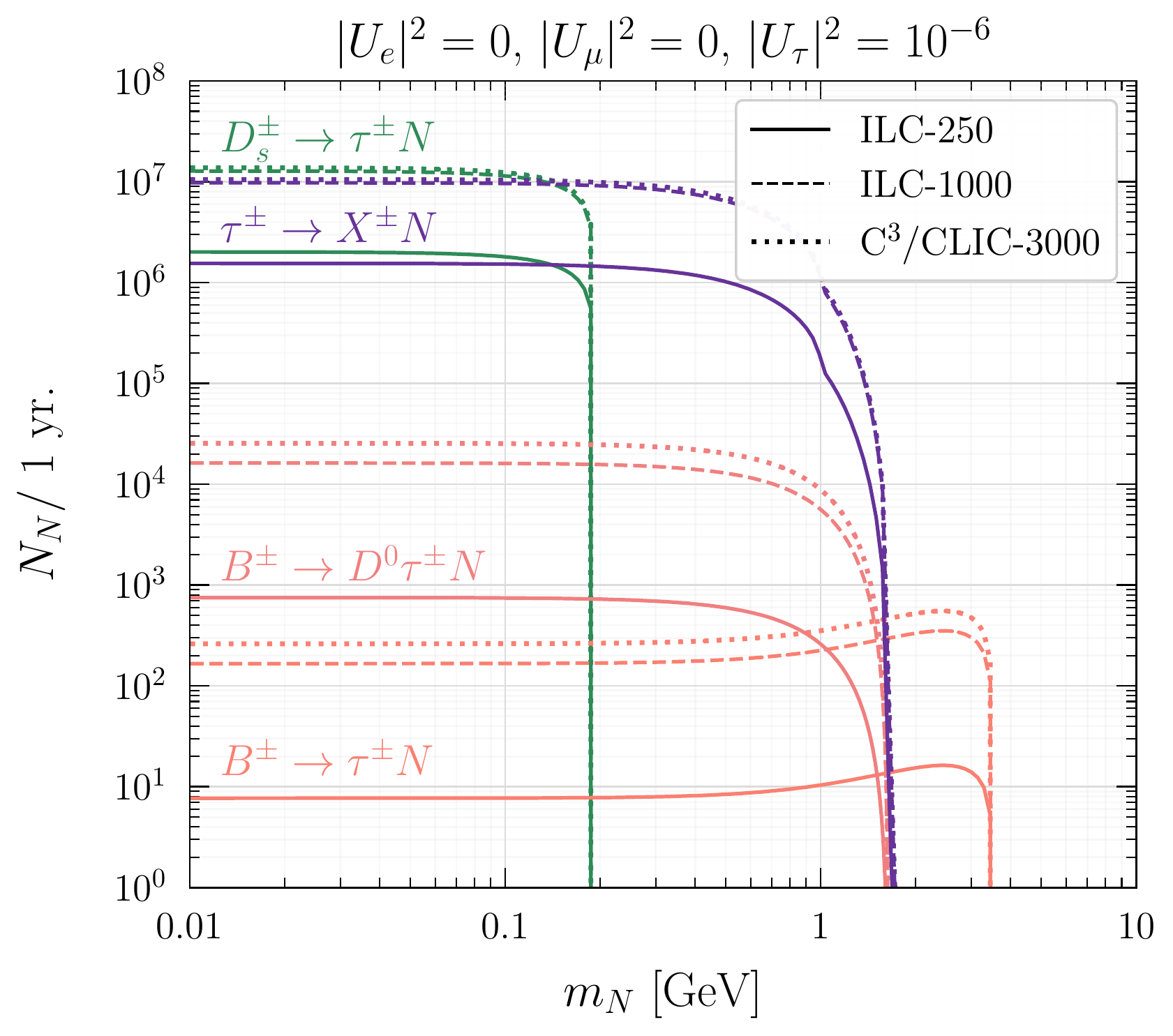}
    \caption{Number of HNLs produced per year at the several experiments, having fixed a mixing angle $|U_i|^2 = 10^{-6}$. \textit{Top left:} Number of electron-mixed HNLs produced from charged-current electron-proton scattering. \textit{Top right:} Number of electron-mixed HNLs produced from the decays of mesons. \textit{Bottom}: Number of muon-mixed (left panel) and $\tau$-mixed (right panel) HNLs
produced from the decays of mesons and $\tau$ leptons. 
\label{fig:HNLProd}  }
\end{figure}

In the top left panel of figure~\ref{fig:HNLProd} we show the number of HNLs produced in charge-current scattering for one year of operation of ILC-250 (solid curve),  ILC-1000 (dashed curve), and C$^3$/CLIC-3000 (dotted curve) assuming a mixing angle of $|U_e|^2 = 10^{-6}$. As we will discuss, this production rate is generally much lower than the one obtained from pion, kaon, and $D$-meson decays. It is, instead, comparable to or larger than the production rate from $B$ meson decays.

For all flavor alignments, HNLs can be copiously produced in two- or three-body decays of mesons, as well as secondary production from the decays of $\tau^\pm$ produced in the decay of $D^\pm_s$ meson if the HNL mixes exclusively with the $\tau$ neutrino. The number of HNLs produced is given by
\begin{equation}\label{NHNL}
N_N = N_M \text{BR}(M \to N + X)\,,
\end{equation}
where $N_M$ is the number of mesons or $\tau^\pm$ produced (see eq. (\ref{eq:Nmeson}) and the subsequent discussion) and BR($M \to N + X$) is the branching ratio for the meson or $\tau^\pm$ decay to a HNL and final states $X$. The full expressions of all branching ratios can be found in \cite{Bondarenko:2018ptm,Shrock:1980vy,Shrock:1980ct,Shrock:1981wq}.

The total number of HNLs produced from meson and $\tau$ decays in an ILC and C$^3$/CLIC beam dump experiment is shown in the top right, bottom left, and bottom right panels of figure~\ref{fig:HNLProd} for electron-mixed, muon-mixed, and $\tau$-mixed HNLs, respectively, assuming one year of operation and a mixing angle of $|U_\alpha|^2 = 10^{-6}$. For all production channels, we sum over positively and negatively charged mesons using their respective production given in table~\ref{tab:Nmesons}.

The production of electron-mixed HNLs from meson decays are typically orders of magnitude larger than the direct production in charged-current scattering, and we expect the charged-current scattering to place weaker constraints for $m_N \lesssim m_B$. At higher masses, charged-current scattering becomes a relevant production mechanism. The production of HNLs at ILC-1000 and C$^3$/CLIC-3000 is generally larger than at ILC-250, especially if it arises from $B$ meson decays, as well as charged current interactions.

\subsection{HNL Decays}\label{subsec:HNLdecay}

\begin{figure*}[t]\label{fig:HNLBR}
    \centering
    \includegraphics[width=0.49\textwidth]{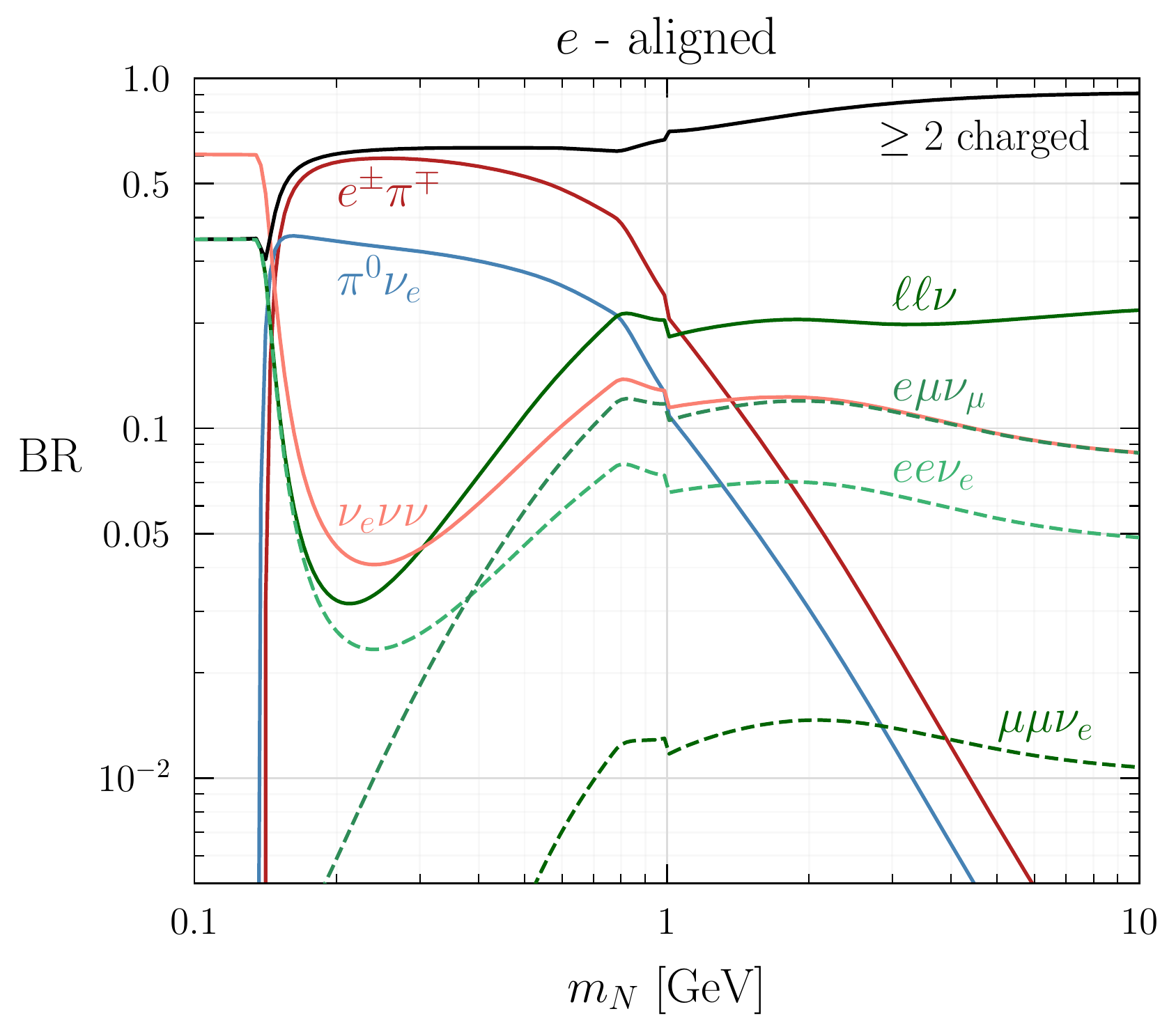}
        \includegraphics[width=0.49\textwidth]{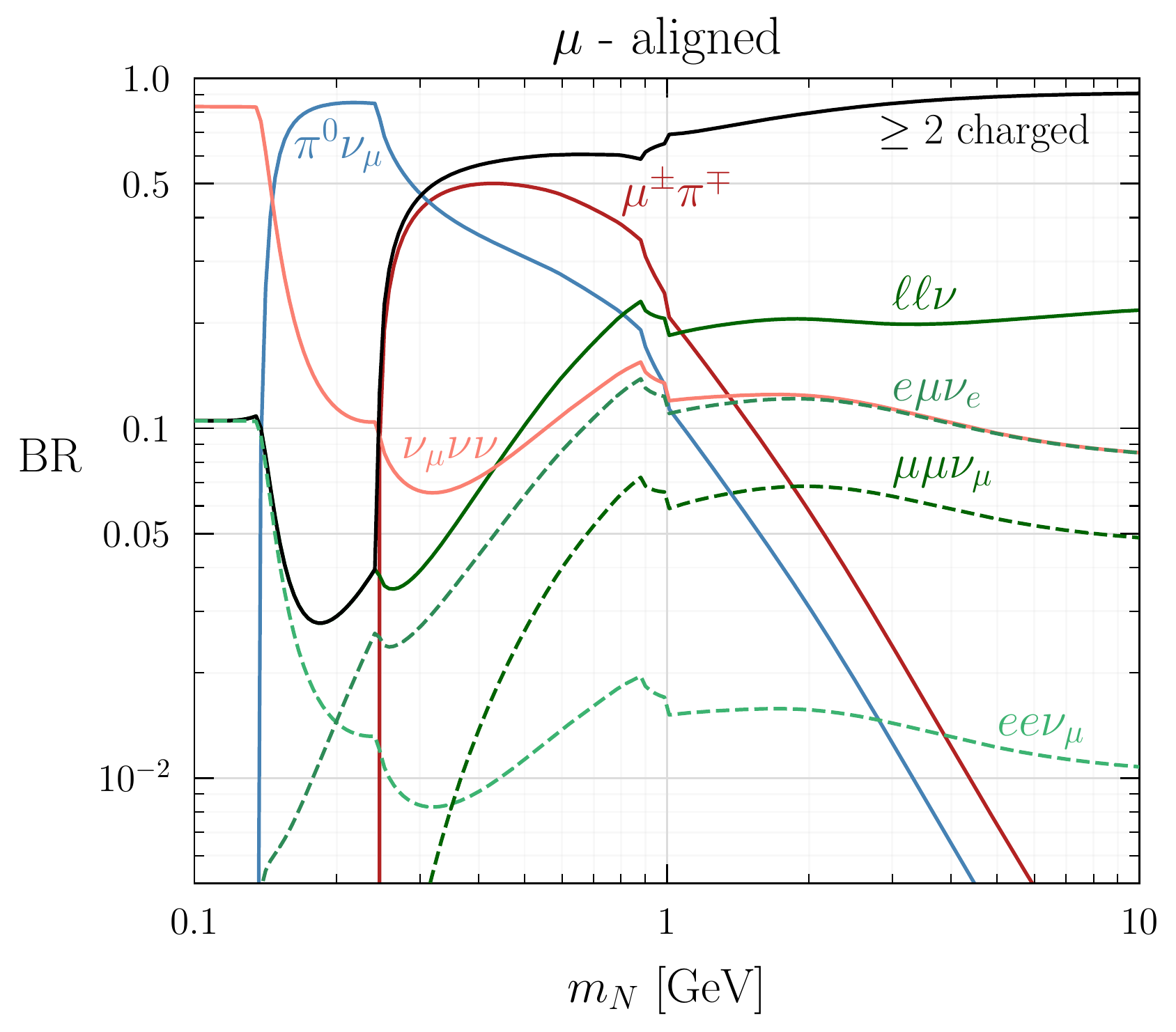}\\
        \includegraphics[width=0.49\textwidth]{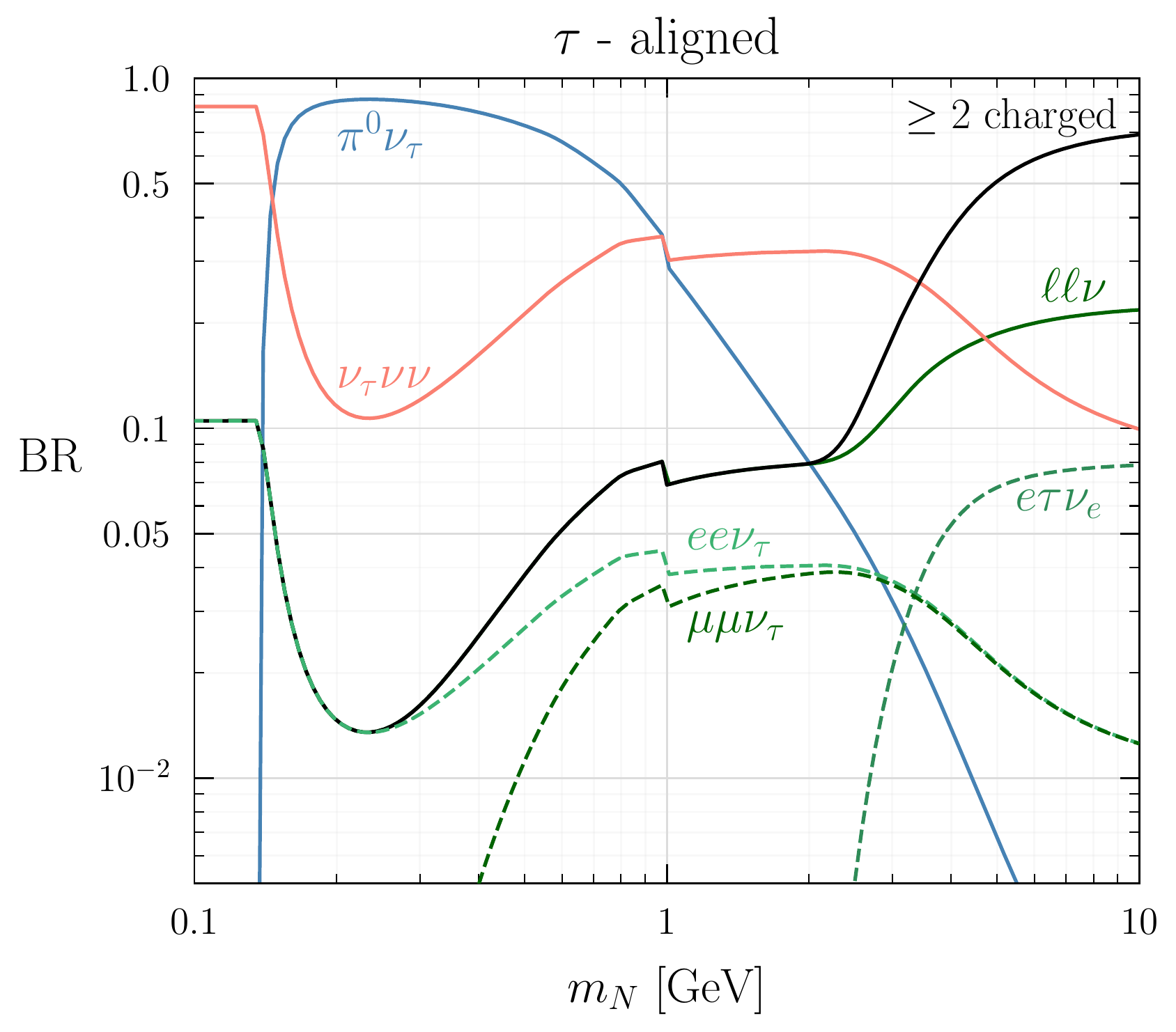}
    \caption{Branching ratios of an electron-mixed (left upper panel), muon-mixed (right upper panel), and $\tau$-mixed (lower panel) HNL into different final states as a function of the HNL mass, $m_N$. The thick black curve represents the sum of the branching ratios into two or more charged particles in the final state, while the solid green curve labeled $\ell\ell\nu$ is the sum of the branching ratios into charged lepton final states.
    }
\end{figure*}

Once produced, HNLs will decay via the weak interactions in eq.~(\ref{eq:Lweak}) into two-body final states, $N \to \ell^\pm M$, $N \to \nu M$, where $M$ is a meson, and three-body final states, $N \to f f^\prime \nu_i$ for final state fermions $f,f^\prime$. The partial widths for the several decays are given in \cite{Gorbunov:2007ak,Atre:2009rg,Bondarenko:2018ptm,Berryman:2017twh}. In figure~\ref{fig:HNLBR} we show the branching ratios as a function of $m_N$ for an electron-mixed (top-left), muon-mixed (top right), and $\tau$-mixed (bottom) HNL. For HNL masses below 1 GeV, the total hadronic decay rate is given by the sum of exclusive decays to mesons, while above 1 GeV we use the inclusive decay to quarks. The several branching ratios do not depend on the mixing angles $|U_i|^2$.

Among the possible decays, the most promising channels are those involving two (or more) charged particles in the final state, as those channels with only neutral particles (e.g., $\pi^0 \nu$) will have larger SM backgrounds and are harder to be detected. To compute the reach on the HNL parameter space, we will take a conservative approach and only consider the HNL decay mode with at least two charged particles and with the largest branching ratio. For an electron- or muon-mixed HNL the most important decay for $m_N \lesssim m_\pi$ is $N \to ee\nu$ (dashed light-green curve), while for $m_\pi\lesssim m_N \lesssim 1$ GeV it is given by $N \to \ell^\pm \pi^\mp~(\ell = e, \mu)$ (red curves). Above $m_N = 1$ GeV, the decay to two charged leptons will be the most relevant, with $N \to e\mu\nu$ being the dominant one. The most relevant decays of the $\tau$-mixed HNL are to two charged leptons, as shown by the green curves in the bottom plot of figure~\ref{fig:HNLBR}.

\subsection{Detector Efficiency and Acceptance}

Given the experimental setup, we can determine the geometric acceptance and efficiency to detect the visibly decaying HNL. We require that the visible final states reach the detector after the HNL has decayed anywhere in the decay volume (see figure \ref{fig:configuration}). For a given decay position, $z$, there will be a requirement on the direction of the visible final states with respect to the beam axis. For each position $z$ there is a maximum angle, $\theta_\text{max}$ given by:
\begin{equation}\label{eq:thetamax}
\tan\theta<  \tan\theta_\text{max} = \frac{r_\text{det}}{z_\text{max} - z},
\end{equation}
where $r_\text{det}$ is the radius of the detector and $z_\text{max} = \ell_\text{dump} + \ell_\text{sh} + \ell_\text{decay} = 131$m is the location of the detector. In fact, visible decay products with an angle $\theta>\theta_\text{max}$ will not reach the detector. Signal events are chosen to be those in which the HNL decays to two charged particles that are within the fiducial decay region enclosed by $z \in (z_\text{min},z_\text{max})$, where $z_\text{min} = \ell_\text{dump} + \ell_\text{sh} = 81$ m, and satisfy eq.~(\ref{eq:thetamax}).

Since the geometric acceptance depends on the particular location of the HNL decay, the total efficiency is given by 
\begin{equation}\label{eq:eff}
\text{eff} = m_N\Gamma \int_{z_\text{min}}^{z_\text{max}} dz \sum_{\text{events}~ \in~\text{geom.}} \frac{e^{-zm_N\Gamma/p_z}}{N_\text{MC} p_z}\,,
\end{equation}
where $m_N$, $\Gamma$, and $p_z$ are the mass, decay width, and $z-$component of the momentum of the HNL, respectively. The sum is over the events that fall within the geometric acceptance of the detector, and $N_\text{MC}$ is the total number of simulated events.

\begin{figure}
    \centering
    \includegraphics[width=\textwidth]{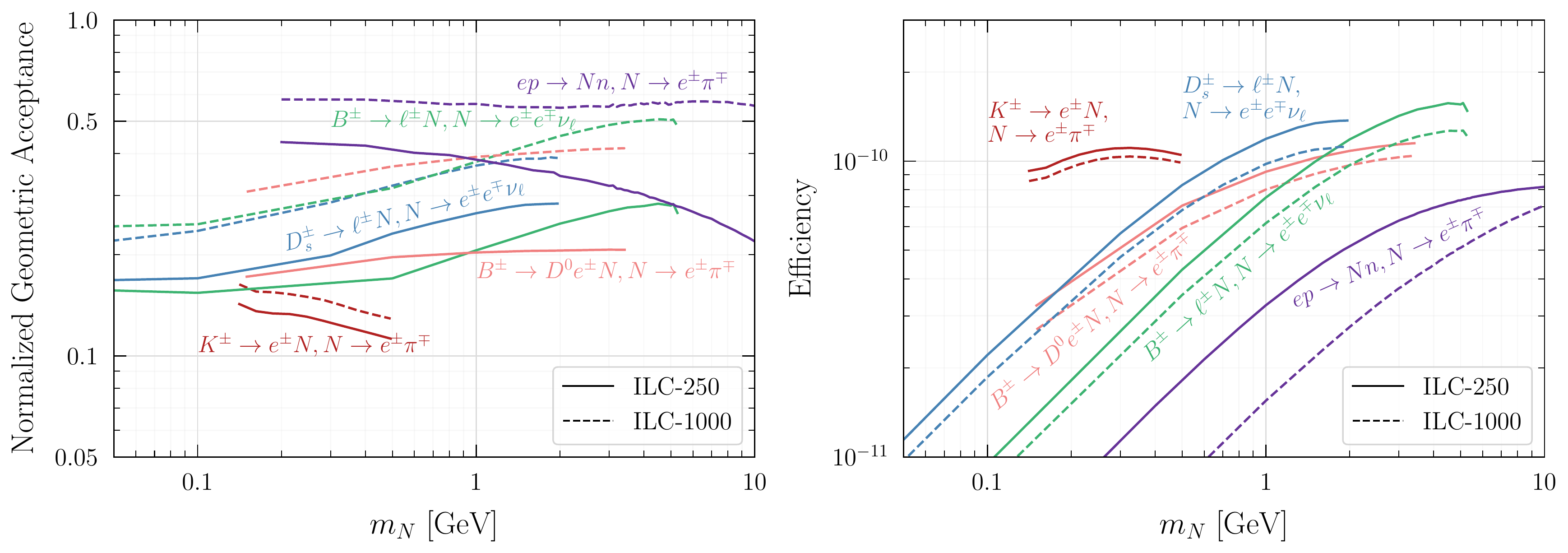}
    \caption{\textit{Left:} Normalized geometric acceptance as a function of the HNL mass in the long lifetime regime (i.e., having fixed the proper lifetime to 30 seconds) for ILC-250 (solid curves) and ILC-1000 (dashed curves).  \textit{Right:} Efficiency as a function of HNL mass. See text for more details. \label{fig:eff}
    }
\end{figure}

The efficiency as a function of the HNL mass is shown in the right panel of figure~\ref{fig:eff} for a few representative HNL production and decay channels at the ILC-250 and ILC-1000. All of these curves apply to an electron-mixed HNL while the blue and green curves apply also to a muon-mixed HNL. In computing these curves, we assume a large proper lifetime of the HNL, 30 seconds, and, therefore, a decay length much larger than the experimental setup. In this limit, we can approximate the efficiency as
\begin{equation}\label{eq:effLong}
\text{eff} \approx m_N \Gamma (z_\text{max} - z_\text{min}) \sum_{\text{events}~ \in~\text{geom.}} \frac{1}{N_\text{MC} p_z}\,.
\end{equation}
As expected, in this regime the efficiency depends linearly on the $|U_\alpha|^2$.
In the right plot of figure~\ref{fig:eff} we can see that the efficiency is smaller for ILC-1000 (dashed curves) compared to ILC-250 (solid curves). This is because, at higher energies, the HNLs produced are more boosted in the forward direction, which gives a smaller probability that the HNL will decay within the decay volume of the experiment (in the long life-time regime). For brevity, we omit the corresponding C$^3$/CLIC-3000 curves from figure~\ref{fig:eff}. The same conclusions hold due to its higher energy: the efficiency is lower than ILC-1000.

The geometric acceptance shows the opposite behavior. In the left panel of  figure~\ref{fig:eff}, we show the acceptance of ILC-250 (solid lines) and ILC-1000 (dashed lines). We compute the  acceptance by taking the ratio between the efficiency in eq.~(\ref{eq:eff}) with or without the requirement to be in geometric acceptance. The geometric acceptance is larger at ILC-1000 compared to ILC-250 which reflects the fact that the final states are more forward, and, therefore, the number of events that satisfy the geometric acceptance requirement is higher. Overall, the geometric acceptance is quite large, thanks to the sizable boost of the HNLs produced in these experiments.

\section{Sensitivity for Heavy 
Neutral Leptons}\label{sec:reach}
\subsection{Sensitivity of Lepton Collider Beam Dump Experiments}
The expected reach of a high-energy electron beam dump experiment can be found using the production, decay, and experimental efficiency information discussed in the previous section. The total number of signal events is given by
\begin{equation}\label{eq:Nsig}
N_\text{sig} = N_N\times \text{BR}(N \to i) \times \text{eff}_i\,,
\end{equation}
where $N_N$ is the number of HNLs produced in a particular production mode, BR($N \to i$) is the branching ratio for $N$ decaying to a final state $i$, and eff$_i$ is the efficiency for detecting the final states. For each mass point, we consider only the HNL decay mode that leads to the best sensitivity, as discussed in section~\ref{subsec:HNLdecay}. To determine the expected reach, we require 10 signal events in the detector.

\begin{figure}
    \centering

    \includegraphics[width=0.495 \textwidth]{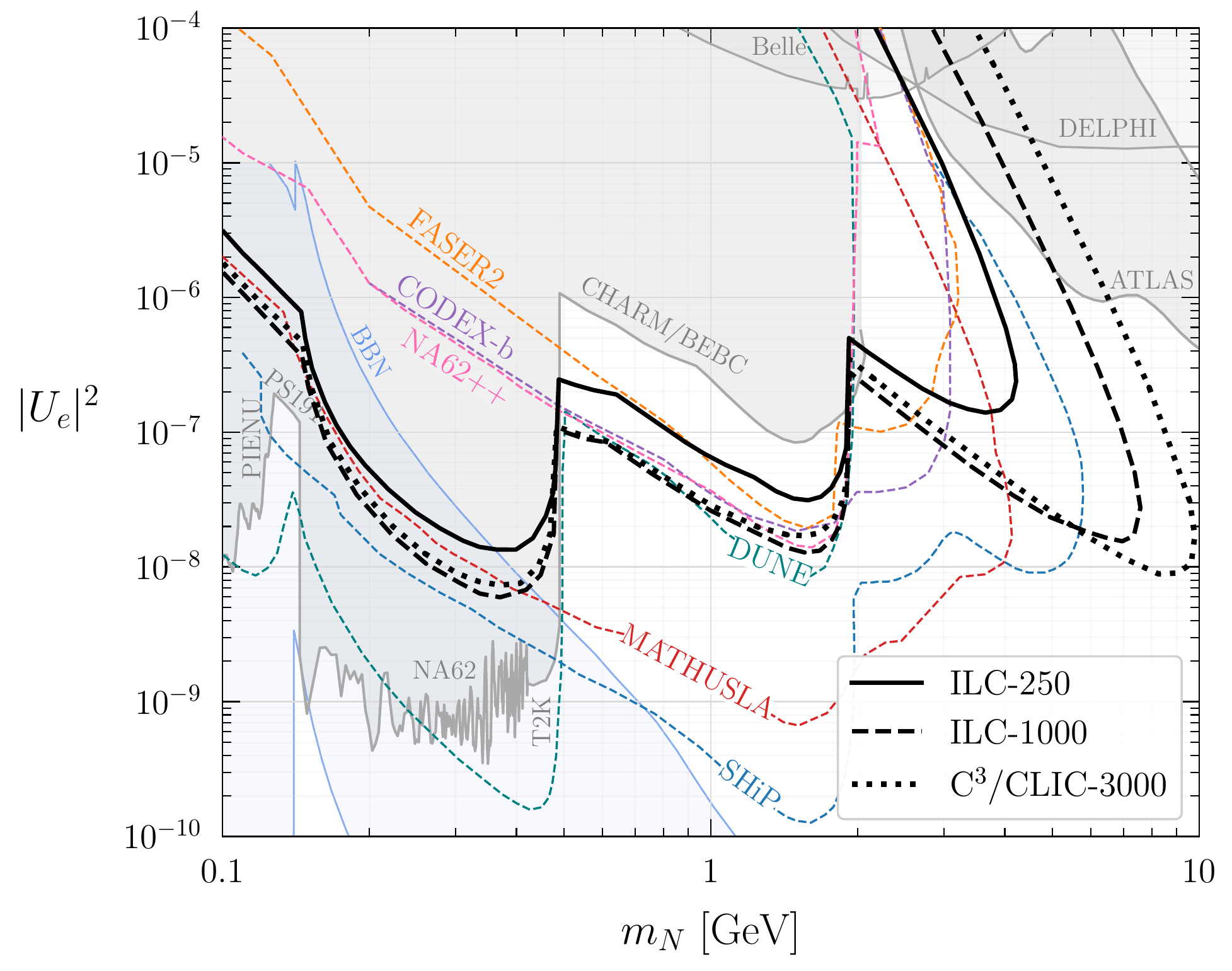}
    \includegraphics[width=0.495\textwidth]{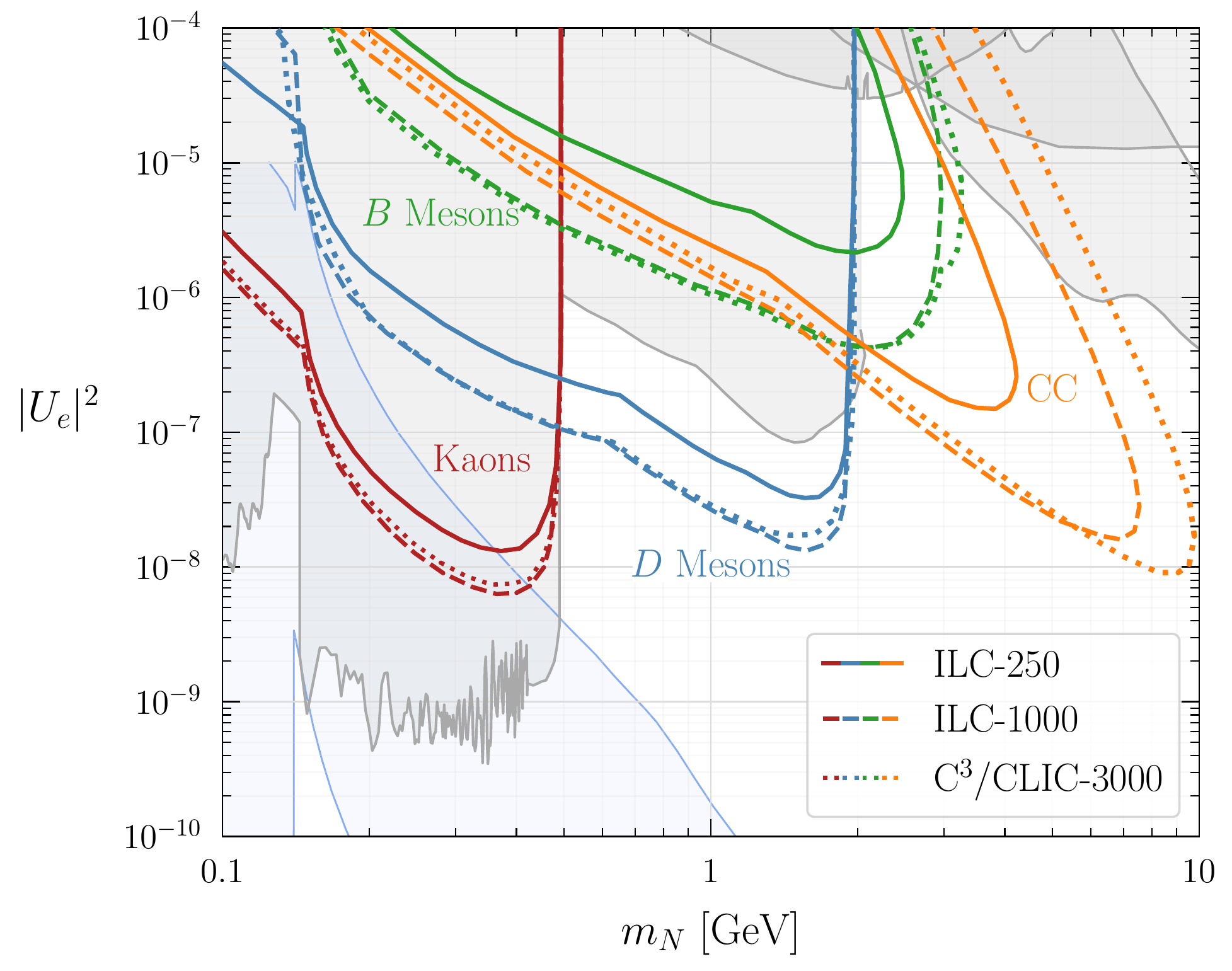}
    \caption{\textit{Left:} The reach of beam dump experiments for electron-mixed HNLs in the $|U_e|^2$ vs $m_N$ plane for ILC-250 (black, solid curve), ILC-1000 (black, dashed curve), and C$^3$/CLIC-3000 (black, dotted curve) assuming one year of operation. The limits are set by requiring 10 signal events in the detector. Existing experimental limits are depicted by the gray shaded region, while constraints from BBN are depicted by the blue shaded region. Limits from proposed experiments are depicted by the different dashed, colored curves. \textit{Right:} Same as the left plot where we decompose the bounds into contributions from the decay of kaons (red curves), $D$ mesons (blue curves), $B$ mesons (green curves), and charged-current scattering (orange curves).
    \label{fig:BoundsElectron}
    }
\end{figure}
\begin{figure}
    \centering
    \includegraphics[width=0.495\textwidth]{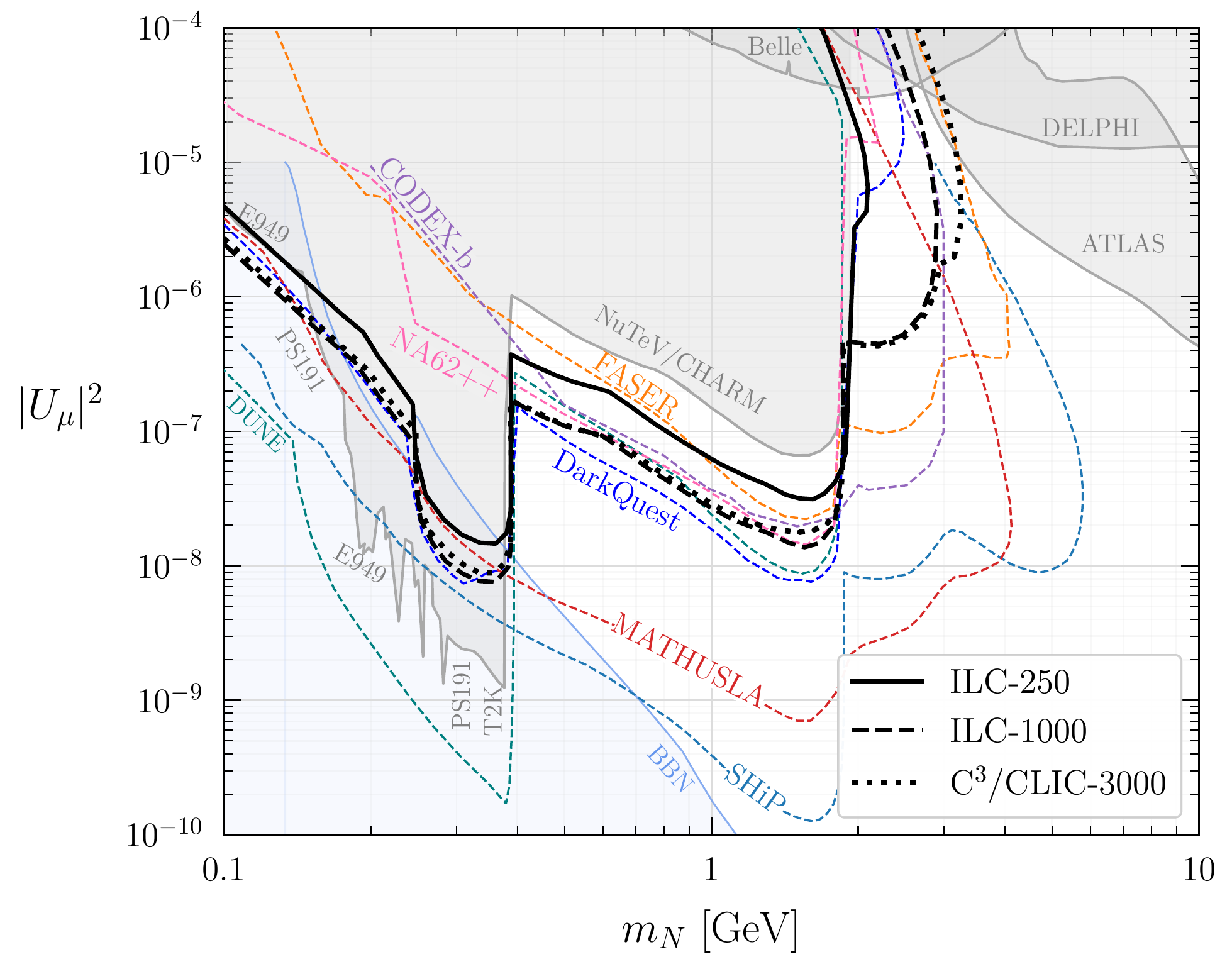}
    \includegraphics[width=0.495\textwidth]{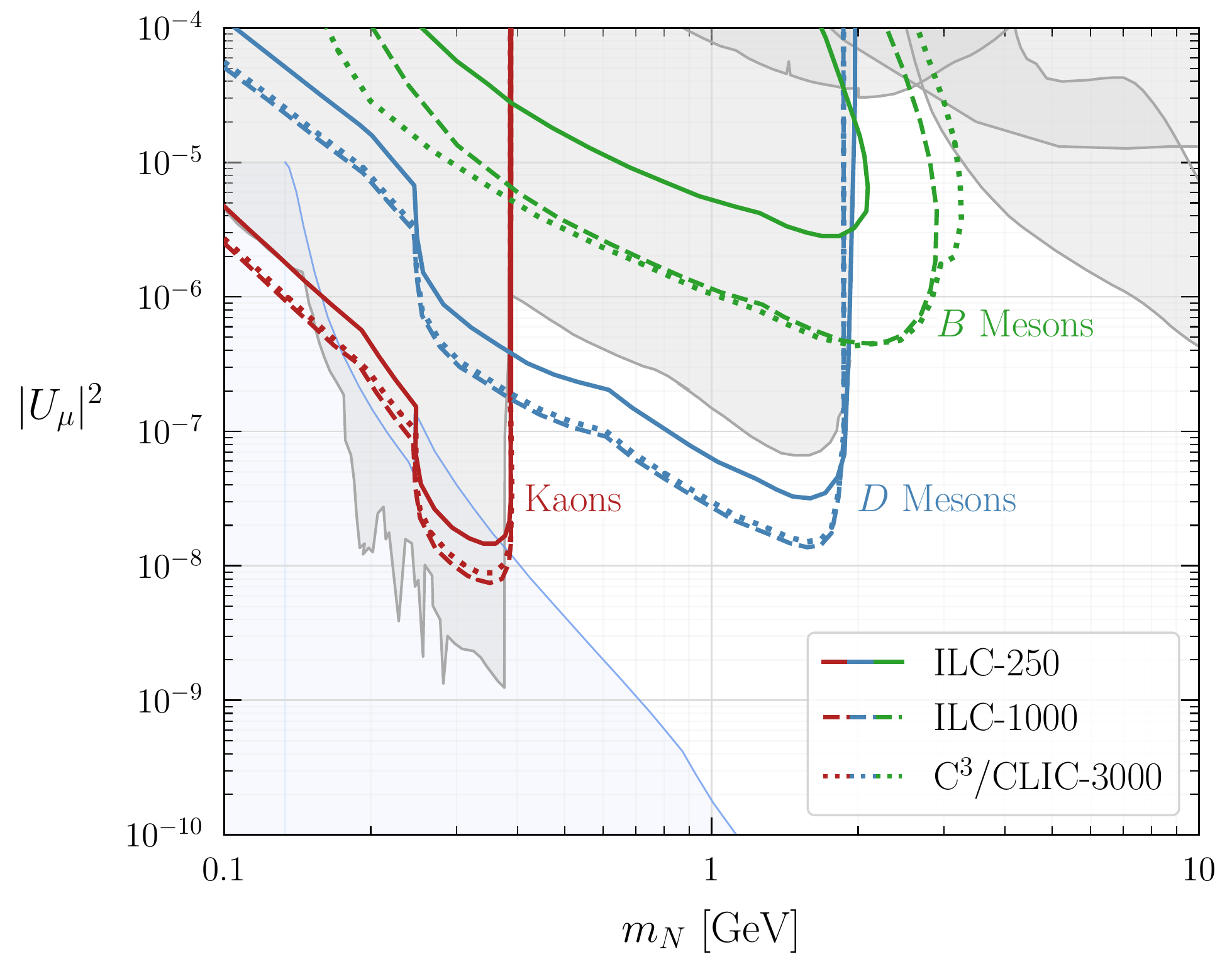}
    \caption{Same as in figure~\ref{fig:BoundsElectron} but for muon-mixed HNLs.
    \label{fig:BoundsMuon}}
\end{figure}
\begin{figure}
    \includegraphics[width=0.495\textwidth]{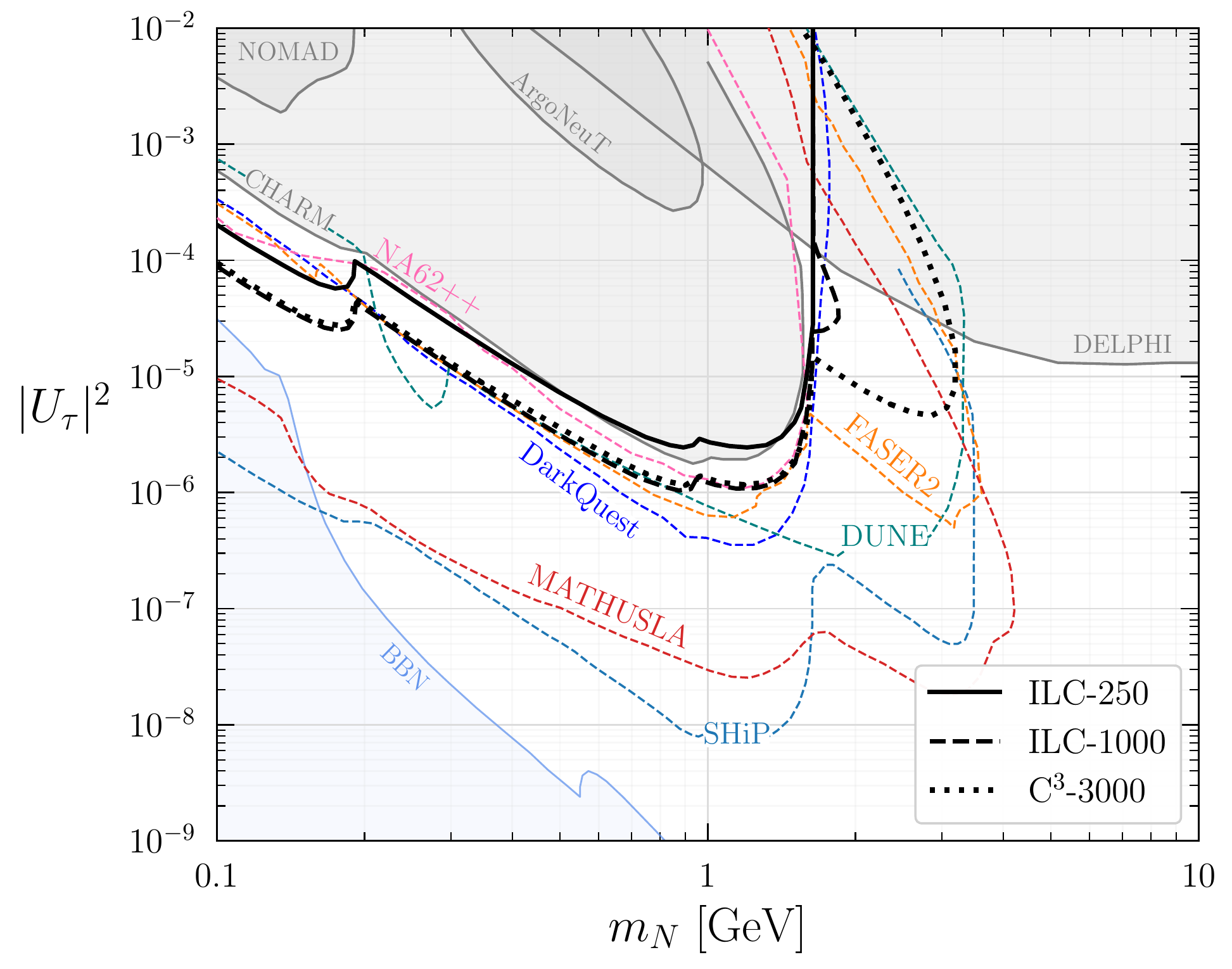}
    \includegraphics[width=0.495\textwidth]{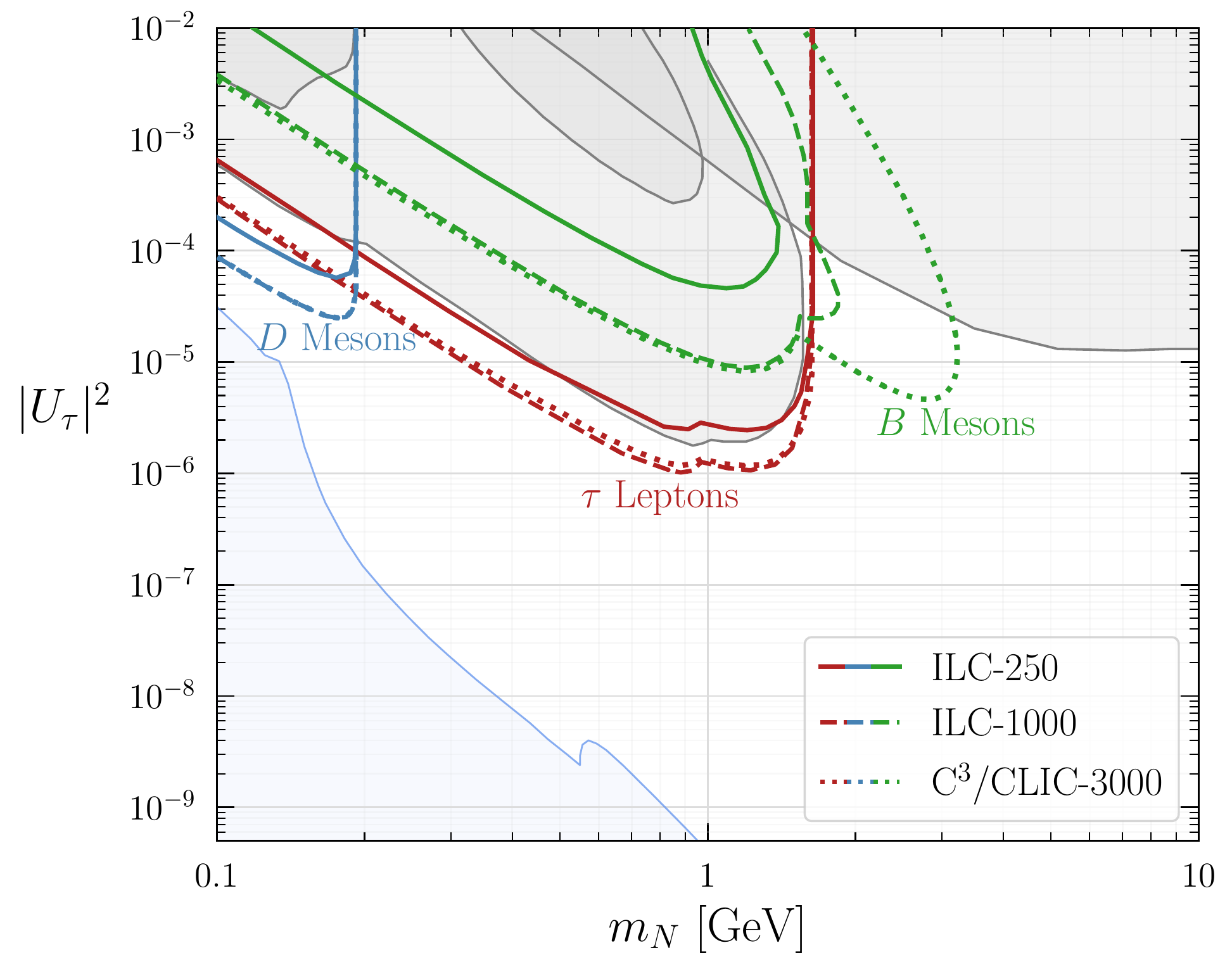}
    \caption{Same as in figure~\ref{fig:BoundsElectron} but for $\tau$-mixed HNLs. \label{fig:BoundsTau}}
\end{figure}

In figures~\ref{fig:BoundsElectron}, \ref{fig:BoundsMuon}, and \ref{fig:BoundsTau}, we show the reach of high-energy electron beam dump experiments for electron-mixed, muon-mixed, and $\tau$-mixed HNLs, respectively, assuming one year of operation. In the left plot of each figure the solid-black, dashed-black, and dotted-black curves represent the reach of the ILC-250, ILC-1000, and C$^3$/CLIC-3000, respectively. In the right plot of figure~\ref{fig:BoundsElectron} and \ref{fig:BoundsMuon}, we decompose the bounds into contributions from the decay of kaons (red curves), $D$ mesons (blue curves), $B$ mesons (green curves), and charged-current scattering (orange curves, only for figure~\ref{fig:BoundsElectron}). For figure~\ref{fig:BoundsTau} we decompose the bounds into contributions from $D$ mesons (blue curves), $\tau^\pm$ (red curves), and $B$ mesons (green curves). 

The gray shaded regions are the leading existing constraints from
PIENU~\cite{PiENu:2015seu,PIENU:2017wbj}, PS191~\cite{Bernardi:1985ny,Bernardi:1987ek}, NA62~\cite{NA62:2020mcv}, CHARM~\cite{CHARM:1985nku}, BEBC~\cite{WA66:1985mfx}, Belle~\cite{Belle:2013ytx}, DELPHI~\cite{Abreu:1996pa}, and ATLAS~\cite{ATLAS:2022atq} for electron-mixed HNLs;
E949~\cite{Artamonov:2014urb}, PS191~\cite{Bernardi:1985ny,Bernardi:1987ek}, T2K~\cite{T2K:2019jwa,Arguelles:2021dqn}, NuTeV~\cite{Vaitaitis:1999wq}, CHARM~\cite{Vilain:1994vg}, Belle~\cite{Belle:2013ytx}, DELPHI~\cite{Abreu:1996pa}, and ATLAS \cite{ATLAS:2022atq} for muon-mixed HNLS; and NOMAD~\cite{NOMAD:2001eyx}, CHARM~\cite{NOMAD:2001eyx,Orloff:2002de}, ArgoNeuT~\cite{ArgoNeuT:2021clc}, and DELPHI~\cite{Abreu:1996pa} for $\tau$-mixed HNLs. The shaded blue regions at small mixing angles are additional constraints from Big Bang nucleosynthesis (BBN) studies~\cite{Boyarsky:2020dzc,Sabti:2020yrt}. HNLs can, in fact, be thermally produced in the early Universe. The decays of these HNLs in the plasma affect the abundance of light elements as well as the effective number of relativistic species, $N_{eff}$. Thus, the HNL parameter space can be constrained by BBN measurements.In all mixing scenarios, the ILC-250, ILC-1000, and C$^3$/CLIC-3000 will cover unexplored regions of parameter space. 

We also compare the sensitivity of the ILC and C$^3$/CLIC beam dump experiments to the sensitivity of several proposed experiments such as CODEX-b~\cite{Aielli:2019ivi}, DarkQuest~\cite{Batell:2020vqn}, DUNE~\cite{Ballett:2019bgd,Berryman:2019dme,Coloma:2020lgy}, FASER~\cite{Kling:2018wct}, MATHUSLA~\cite{Curtin:2018mvb,MATHUSLA:2020uve}, NA62~\cite{Drewes:2018gkc}, and SHiP~\cite{Gninenko:2012anz,Bonivento:2013jag,Alekhin:2015byh,SHiP:2018xqw,Gorbunov:2020rjx} (see the colored dashed lines in the left plot of the figures). With only one year of operation, the ILC and C$^3$/CLIC can lead to a comparable or even better sensitivity as FASER, NA62, DarkQuest, CODEX-b, and DUNE.

For all HNL mixings, the sensitivity on $|U_\alpha|^2$ for $m_N\lesssim m_{D^\pm_s},m_\tau$ improves only by a factor of a few when going from the ILC-250 to the ILC-1000 due the an increase in the number of mesons produced per electron-on-target. However, the ILC-1000 bound is not a simple rescaling with the number of mesons since the efficiency of ILC-1000 is lower compared to ILC-250, as shown in the right plot of figure~\ref{fig:eff}. The HNLs are more boosted in the forward direction, resulting in a smaller probability that the HNL will decay in the decay volume. C$^3$/CLIC-3000 places similar limits as ILC-1000.

For $m_N > m_{D^\pm_s}$  using a higher energy electron beam leads to significant improvements in the bounds from HNL production in $B$ decays. Approximately 10 times more $B$ mesons are produced at ILC-1000 and C$^3$/CLIC-3000 compared to ILC-250. This helps increase the sensitivity for
small mixing angles, $|U_{e,\mu}|^2 \lesssim 10^{-5}$  and $|U_\tau|^2 \lesssim 10^{-4}$. For larger mixing angles, $|U_{e,\mu}|^2 \gtrsim 10^{-5}$  and $|U_\tau|^2 \gtrsim 10^{-4}$, there are significant gains in sensitivity when going to higher energies.  In this regime, the bound is determined by the location of the beginning of the decay volume: if the mixing angle is too large, the HNL will decay before it enters the decay volume. Then, for higher energy electron beams, the HNLswill have a larger boost factor and, therefore, a longer lifetime. Therefore, a larger mixing angle will be needed for the HNL to decay at the beginning of the decay volume.

Interestingly, the charged-current production of electron-mixed HNLs places the dominant sensitivity for $m_N >m_{D^\pm_s}$, and can significantly exceed the sensitivity from  $B$-meson decays due to the much lower $B$ meson production rate (see the right plot of figure~\ref{fig:BoundsElectron}). Thanks to the charged-current production, HNLs with a mass up to $\sim 4,7,10$ GeV could be probed at ILC-250, ILC-1000, and C$^3$/CLIC-3000, respectively. This is a unique advantage of electron beam dump experiments over proton beam dump experiments where HNLs are mainly produced via meson decays and are limited by the heaviest mesons that can be produced at a given center-of-mass energy. 
Furthermore, contrary to the production from kaon and $D$ meson decays, increasing the beam energy leads to a significant gain in sensitivity from charged-current production. This can be understood from the upper left plot of figure~\ref{fig:HNLProd}.  ILC-1000 and C$^3$/CLIC-3000 can provide the leading sensitivity for electron-mixed HNLs in the 5 -- 10 GeV mass range, providing a stronger reach than proton beam dump experiments.


\subsection{Modifying the Experimental Setup}\label{sec:mod}

\begin{figure*}[t]
    \centering
    \includegraphics[width=0.495\textwidth]{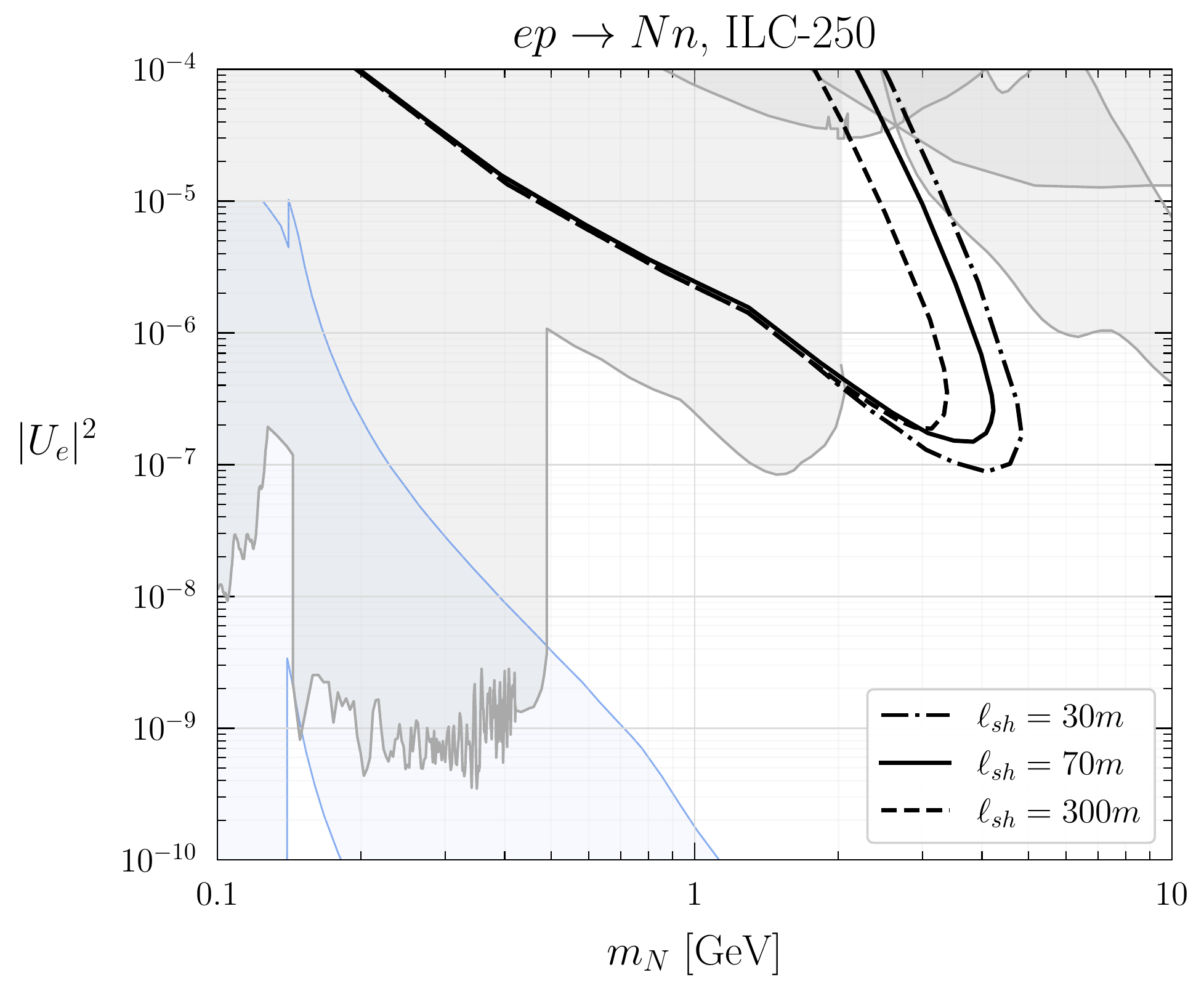}
    \caption{
    Variation of the bound on a electron-mixed HNL parameter space for different lengths of the muon shield $\ell_{\rm sh}$ (fixing the distance to the dump $\ell_{\rm dump}$, the decay length $\ell_{\rm decay}$, and the energy of the beam, 125 GeV). For the purpose of the plot, we only consider the charged-current production of electron-mixed HNLs for ILC-250.
    \label{fig:BoundsVary}}
\end{figure*}

In the previous section, we have analyzed the sensitivity of an ILC and C$^3$/CLIC beam dump experiment using the ILC beam configuration described in section~\ref{Sec:Setup}. We have seen that increasing the electron beam energy leads to gains in the sensitivity for both small mixing angles (lower part of the curves) and for large mixing angles at high HNL masses (upper part of sensitivity curves). Here we show the impact of different parameters of the experimental setup in the reach of HNL parameter space.

An adjustable parameter of the experimental set up is the length of the muon shield, $\ell_\text{sh}$. Varying the length of the shield does not affect the bounds at small values of the mixing angle (the lower part of the bounds). 
In fact, in this regime the efficiency varies linearly with the size of decay length (see eq.~(\ref{eq:effLong})) which we keep fixed. On the other hand, varying the length of the shield will change the sensitivity for HNLs that are relatively short lived (the upper part of the bounds). As a result, we only consider the direct production of electron-mixed HNLs from charged-current scattering, which has the best sensitivity in that region of parameter space.

We consider $\ell_\text{sh} =$ 30 m, 70 m (i.e. the configuration described in section~\ref{Sec:Setup}), and 300 m as benchmarks, while keeping all other parameters the same as the nominal beam dump configuration (see figure~\ref{fig:configuration}). The resulting bounds for ILC-250 are shown in figure~\ref{fig:BoundsVary} for the example of an electron-mixed HNL. The dot-dashed, solid, and dashed curves are for a shield length of 30~m, 70~m, and 300~m, respectively. As expected, changing the length of the shield only affects large values of the mixing angle, and we can see that a shorter shield length improves the sensitivity while a longer shield length decreases the sensitivity, compared to the nominal beam dump set up with $\ell_\text{sh} =$ 70 m. This seems to hint that a small muon shield would be beneficial to obtain a better reach with a lower energy electron beam, potentially closing the gap in parameter space between 2--5 GeV.  However, as we discuss in section~\ref{Sec:BackgroundAndOther}, a muon shield that is too thin can lead to sizable SM backgrounds, depleting, therefore, the overall reach.

\subsection{Possible Backgrounds}\label{Sec:BackgroundAndOther}

Long-lived mesons, especially $K_L^0$, could be background to the searches proposed in this paper. With the experimental setup adopted in our paper, $\mathcal O(10^{16}-10^{17})$ $K_L^0$ will be produced in the first radiation length of the dump. The water dump will be able to absorb some of these kaons, reducing the number to $\mathcal O(10^{12}-10^{13})$. As discussed in \cite{Sakaki:2020mqb}, it is assumed that the inner parts of the muon shield are made of lead. Considering the pion and kaon interaction length in lead (20 cm), we find that roughly 6 meters of muon shield will be able to reduce the kaon rates to negligible values. In all our estimates of the reach on HNL parameter space, we assume $l_{\rm{sh}}\gtrsim 30$ m, and therefore the $K_L^0$ background should be negligible.

In addition, a large flux of muons will be produced. Approximately $\mathcal O(10^{18})$ muons will be produced in the first radiation length of the dump.
The average energies of these muons are roughly 10 GeV, 20 GeV and 30 GeV at ILC-250, ILC-1000, and C$^3$/CLIC-3000, respectively. The muon energy distribution is broad, with sizable populated tails reaching 100 GeV, even at the ILC-250. The muons will lose only a few GeV energy in the water dump. In the lead muon shield, the muons will loose approximately 1 GeV/m. These considerations show that additional veto counters placed after the muon shield would be needed to reduce the muon background to a negligible rate for a muon shield of length $\ell_\text{sh}$ = 30m or 70m.

\section{Summary}\label{sec:conclusion}
We have investigated the sensitivity on the heavy neutral lepton parameter space of beam dump experiments that utilize the high energy electron beam of future $e^+e^-$ colliders. We focused on the 250 GeV and 1000 GeV ILC stages, as well as on the 3 TeV stages of C$^3$ and CLIC experiments. In fact, linear colliders offer a better opportunity for high energy electron beams experiments compared to circular colliders thanks to their much larger number of electrons dumped per second. 

We pointed out that HNLs can be copiously produced in the beam dump of high energy, high intensity $e^+e^-$ colliders either from meson decays, or from charged current electron-proton scattering. These production mechanisms are complementary to the production processes that occur in the main $e^+e^-$ collider.
Once produced the HNLs can travel a macroscopic distance before decaying into SM
particles, that can be detected if a suitable detector is installed behind the beam
dump. These signatures are again complementary to the signatures that can be looked for at the main detectors of the $e^+e^-$ colliders, and, depending on the specific experimental apparatus, could have a very small or negligible SM background. 

We showed that, thanks to their unprecedented energies and intensities, these electron beam dump experiments
will probe a large range of HNL parameter space for HNL masses below $\sim 10$ GeV, not yet probed by past searches. These experiments will be complementary
to other proposed experiments such as proton beam dump experiments,
neutrino experiments, and LHC auxiliary detectors. Particularly, in the case of an electron-mixed HNL, they will be able to reach higher masses thanks to the charged current electron-proton scattering process. 

\bigskip

\noindent
{\bf Note added --}
while working on this project, we learned about a similar work in progress authored by M. Nojiri, Y. Sakaki, K. Tobioka, and D. Ueda \cite{Nojiri:2022xqn}. We coordinated the arXiv submission to be on the same day.


\section*{Acknowledgements}
We are grateful to I. Helenius and P. Ilten for help with simulating meson production at electron beam dump experiments with \texttt{Pythia}, and  J. Issacson for helpful discussions about \texttt{Pythia}. We also thank P. Coloma, I. Jaegle, and K. Kelly for valuable discussions regarding the meson productions at LBNF, electron-ion colliders, and HNLs, respectively. We thank E. Nanni, M. Peskin, and C. Vernieri for helpful discusson about the $\rm C^3$ configuration.
The research of SG and PG is supported in part by the NSF CAREER grant PHY-1915852.
YDT is supported by U.S. National Science Foundation (NSF) Theoretical Physics Program, Grant PHY-1915005. The work of DT is supported by the Arthur B. McDonald Canadian Astroparticle Physics Research Institute. Part of this manuscript has been authored by Fermi Research Alliance, LLC under Contract No. DE-AC02-07CH11359 with the U.S. Department of Energy, Office of Science, Office of High Energy Physics. YDT thank the KITP at the University of California, Santa Barbara, for its program ``Neutrinos as a Portal to New Physics and Astrophysics'', supported in part by the National Science Foundation under Grant No. NSF PHY-1748958. This work was partially performed at the Aspen Center for Physics, which is supported by the National Science Foundation grant PHY-1607611.

\appendix

\section{Simulation Details}\label{app:sim}

\subsection{Meson Productions}

 \setlength{\tabcolsep}{0.65em}
{\renewcommand{\arraystretch}{1.1} 
\begin{table*}[t] 
\centering
\begin{tabular}{  | c | c | c | c |}
\hline
    Meson Type  & 125 GeV  &  500 GeV & 1.5 TeV   \\ \hline
	$\pi^+$ & 2.95 &  3.43 & 3.87 \\ \hline
	$\pi^-$ & 2.65 &  3.13& 3.58\\\hline	
	$K^+$ & 0.26 &   0.31  & 0.37 \\ \hline	
	$K^-$ & 0.21 & 0.27& 0.32\\ \hline
	$K^0_L$ & 0.22 & 0.27 & 0.32\\ \hline
	$D^+$  &  $3.81\times 10^{-4}$&  $5.76\times 10^{-4}$ & $8.19 \times 10^{-4}$  \\ \hline	
	$D^-$ & $4.11\times 10^{-4}$ &  $6.16\times10^{-4}$ & $8.48\times10^{-4}$ \\ \hline
	$D_s^+$& $1.08\times 10^{-4}$  &  $1.79\times10^{-4}$ & $2.52\times10^{-4}$ \\ \hline	
    $D_s^-$& $1.18\times 10^{-4}$  &$1.93\times10^{-4}$ & $2.58\times10^{-4}$ \\ \hline	
	$B^+$ & $4.60\times10^{-7}$   &$1.80\times 10^{-6}$ & $3.30\times 10^{-6}$  \\ \hline
	$B^-$ & $1.30\times10^{-7}$&$1.50\times10^{-6}$& $3.30\times 10^{-6}$ \\ \hline
  \end{tabular}
\caption{
The number of mesons produced per generated Monte Carlo soft QCD event ($N_{M,\text{MC}}/N_\text{MC}^\text{tot}$) at the three different beam energies. To obtain the number of mesons per EOT, we have to multiply these numbers by 
$\sigma_\text{SoftQCD}/\sigma_{eN}$, as described in the text. 
Note that here we are quoting the electron beam energy instead of the center-of-mass energy of the corresponding collider.
}\label{tab:mesonfrac}
\end{table*}}

In electron-proton collisions, mesons are mainly produced when a photon is emitted from the incoming electron and interacts with partons in the proton. The photon can interact directly with quarks inside the proton or it can fluctuate into a hadronic state and interact with quarks and gluons in the proton. The total cross-section for photon-hadron collisions is dominated by soft QCD processes. To simulate the production of mesons, we use \texttt{Pythia~8.3}~\cite{Sjostrand:2006za,Sjostrand:2014zea} using the \texttt{SoftQCD} event class which generates all processes having significant contributions to the total hadronic cross-section. We point out that we are pushing Pythia beyond its nominal operation, so there will be some significant uncertainty associated with our calculations.

To determine the total production rate of mesons we run \texttt{Pythia~8.3} using the flags \texttt{SoftQCD:all = on}, \texttt{PDF:beamA2gamma = on}, and \texttt{Photon:ProcessType = 0}. 
The result of this simulation is given in table~\ref{tab:mesonfrac} where we show the number of mesons produced per generated soft QCD Monte Carlo event for an incoming electron beam with an energy of 125 GeV, 500 GeV, and 1.5 TeV. The total number of mesons produced per EOT is determined by
\begin{equation}
   n_M \equiv \frac{N_M}{\text{EOT}} = \frac{\sigma_\text{SoftQCD}}{\sigma_{eN}} \frac{N_{M,\text{MC}}}{N_\text{MC}^\text{tot}}\,,
\end{equation}\label{eq:nMeson}
where $\sigma_{eN}$ is the total electron-nucleus cross section ($\sim 46$ mb) and $\sigma_\text{SoftQCD}$ is the total hadronic cross section estimated by Pythia ($\sim 2 \times 10^{-3}$ mb, $\sim 7 \times 10^{-3}$ mb, $\sim 10^{-2}$ mb for the 125 GeV, 500 GeV, and 1500 GeV beam energy, respectively). $N_{M,\text{MC}}$ is the total number of mesons produced by Pythia and $N_\text{MC}^\text{tot}$ is the total number of Monte Carlo events that were simulated.  $n_M$ is the input for eq.~(\ref{NHNL}) when we calculate the number of HNLs produced from meson decays.

To determine the geometric acceptance and efficiency of the electron beam dump experiment, we also need to know the four-momenta of the mesons produced in the electron collision with the beam dump. To generate a large sample of mesons, we use different settings for the QCD production as follows:
\begin{itemize}
\item For $\pi^\pm$, $K^\pm$: \texttt{"SoftQCD:all = on"} is used.
\item For $D^\pm, D_s^\pm$ : \texttt{"HardQCD:all = on"} and \texttt{"PhotonParton:all=on"} are used.
\item For $B^\pm$: \texttt{"HardQCD:qqbar2bbbar = on"} and \texttt{"PhotonParton:ggm2bbar = on"} are used.
\end{itemize}

\begin{figure*}[t]
    \centering
    \includegraphics[width=0.49\textwidth]{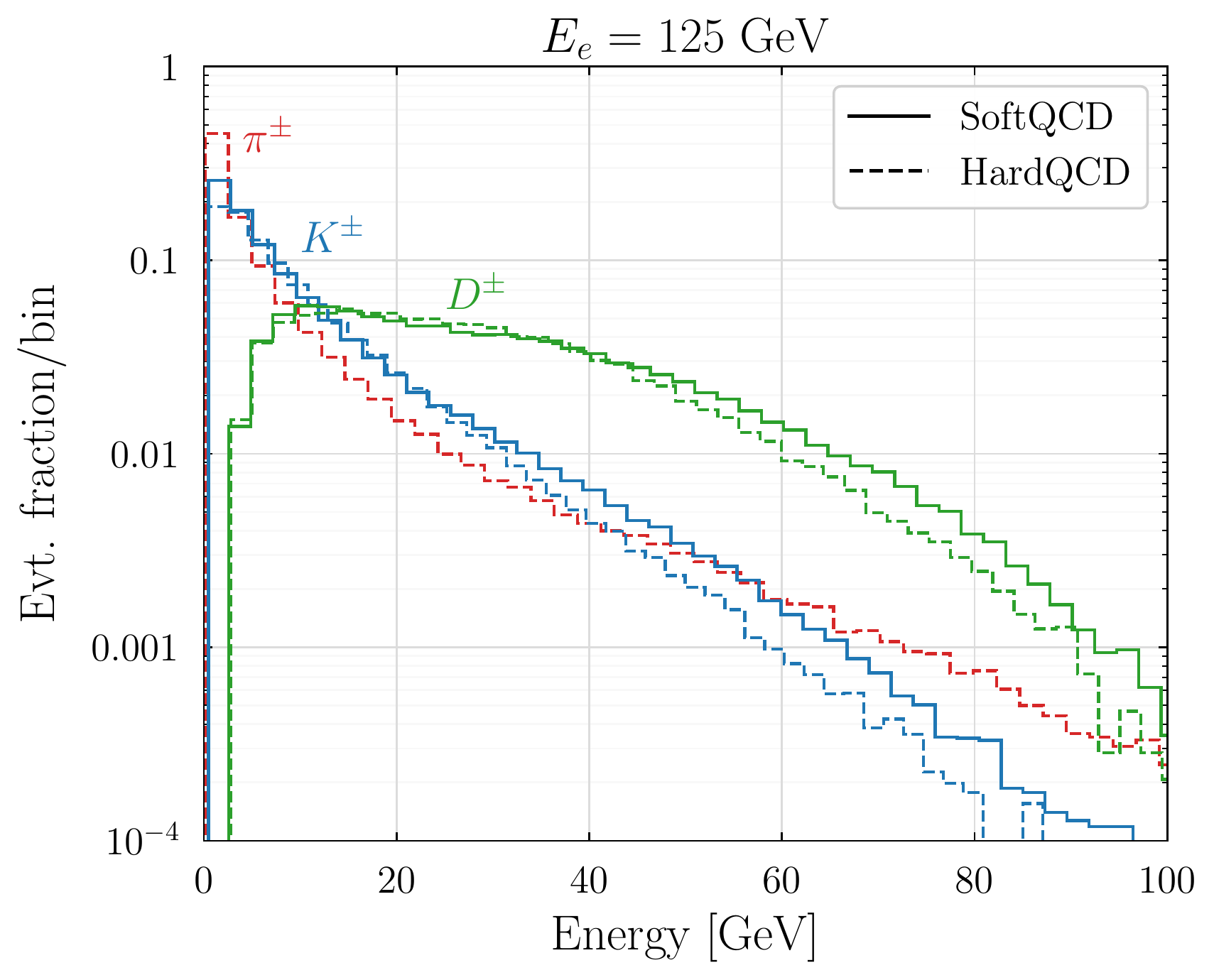}
    \includegraphics[width=0.49\textwidth]{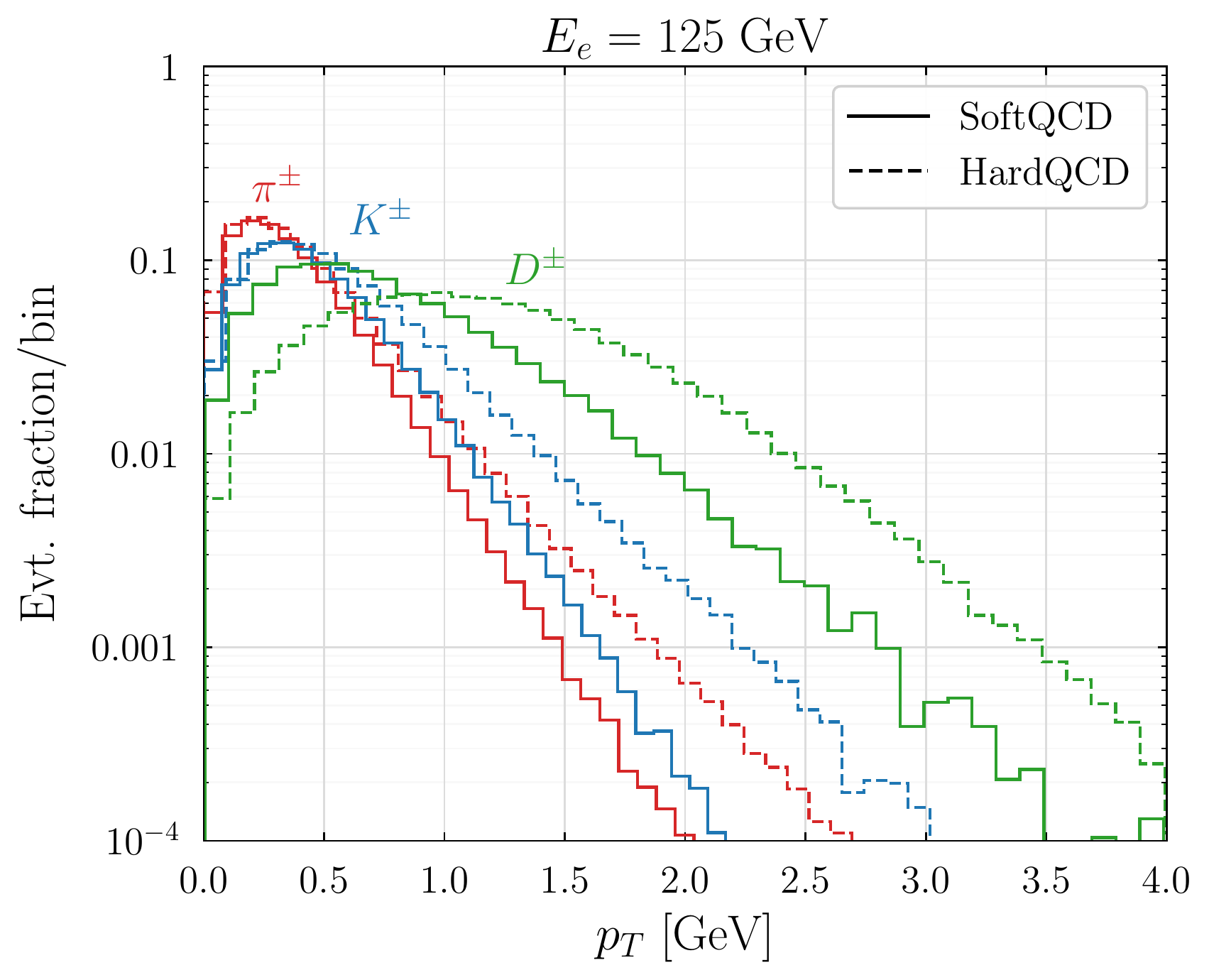}
    \caption{Energy (left) and $p_T$ (right) distributions of mesons produced at an ILC beam dump experiment with an electron beam with energy $E_e =$ 125 GeV hitting a proton at rest. Events are generated with \texttt{Pythia} using the \texttt{SoftQCD} event class (solid curves) and the \texttt{HardQCD} event class (dashed curves).
    The distributions are normalized such that the total bin count sums to 1.
\label{fig:pTcompare}}
\end{figure*}

The \texttt{HardQCD} and \texttt{PhotonParton} settings were used for $D^\pm_{(s)}$ and $B^\pm$ mesons because with the \texttt{SoftQCD} event class the number of these mesons produced per electron on target is very small (cf. table~\ref{tab:mesonfrac}). While \texttt{SoftQCD} will give a more realistic meson spectrum, generating a reasonable number of events to analyze will be computationally very time-consuming, e.g., $\mathcal{O}(10^{11})$ events will need to be generated to produce $\sim$10k $B$ mesons. The number of mesons produced per electron on target using the \texttt{HardQCD} event class is significantly larger. Therefore, we use it to obtain a large sample $D$ and $B$ mesons.
 The \texttt{HardQCD} event class generates mesons with slightly higher $p_T$ and larger energy, as shown in figure~\ref{fig:pTcompare}. This leads to a slightly lower geometric acceptance. However, we have checked that the limits on the HNL parameter space do not vary appreciably if one uses the spectrum of mesons produced with  \texttt{SoftQCD} or \texttt{HardQCD} event class. 
 
\subsection{Three-Body Decays of HNLs}
To determine the sensitivity to the HNL parameter space requires simulating the decays of HNLs to two- and three-body final states and using the resulting four-momenta to determine the geometric acceptance of the detector. For two-body HNL decays (e.g., $N \to \ell^\pm \pi^\mp$), this can be done analytically, starting from the rest frame of the HNL and boosting into the lab frame. For the three-body HNL decays to charged leptons, we use the publicly available code \texttt{muBHNL} \cite{Kelly:2021xbv,deGouvea:2021ual,Kelly:muBHNL}, which uses the differential decay distributions of the HNL to generate a weighted sample of final state leptons. Events are generated in the rest frame of the HNL. They are then boosted to the lab frame using the 4-momenta of mesons generated by Pythia. We then use eq.~(\ref{eq:eff}) to compute the efficiency and eq.~(\ref{eq:Nsig}) to calculate the reach on the HNL parameter space.

\bibliography{references}
\end{document}